\documentclass[preprint]{aastex}
\usepackage{natbib}
\usepackage{latexsym}		
\usepackage{graphicx}		
\usepackage{rotating}           
\usepackage{hyperref}

\shorttitle{Ten Lensing Clusters}
\shortauthors{Wiesner et al.}

\begin{document}

\title{\bf{The Sloan Bright Arcs Survey:  Ten Strong Gravitational Lensing Clusters and Evidence of Overconcentration}}

\author{Matthew P. Wiesner\altaffilmark{1}, Huan Lin\altaffilmark{2}, Sahar S. Allam\altaffilmark{2}, James Annis\altaffilmark{2}, Elizabeth J. Buckley-Geer\altaffilmark{2}, H. Thomas Diehl\altaffilmark{2}, Donna Kubik\altaffilmark{2}, Jeffrey M. Kubo\altaffilmark{2}, Douglas Tucker\altaffilmark{2}}
\altaffiltext{1}{Department of Physics, Northern Illinois University, DeKalb, IL, 60115}
\altaffiltext{2}{Center for Particle Astrophysics, Fermi National Accelerator Laboratory, P.O. Box 500,  Batavia, IL, 60510}

\begin{abstract}
We describe ten strong lensing galaxy clusters of redshift $0.26 \leq z \leq 0.56$ that were found in the Sloan Digital Sky Survey.  We present measurements of richness ($N_{200}$), mass ($M_{200}$) and velocity dispersion for the clusters.  We find that in order to use the mass-richness relation from \citet{Johnston07}, which was established at mean redshift of $0.25$, it is necessary to scale measured richness values up by $1.47$.  Using this scaling, we find richness values for these clusters to be in the range of $22 \leq N_{200} \leq 317$ and mass values to be in the range of $1\times10^{14} h^{-1}M_{\sun} \leq M_{200} \leq 30\times10^{14} h^{-1}M_{\sun}$.    We also present measurements of Einstein radius, mass and velocity dispersion for the lensing systems.  The Einstein radii ($\theta_E$) are all relatively small, with $5.4\arcsec \leq \theta_E \leq 13\arcsec$.  Finally we consider if there is evidence that our clusters are more concentrated than $\Lambda$CDM would predict.  We find that six of our clusters do not show evidence of overconcentration, while four of our clusters do.  We note a correlation between overconcentration and mass, as the four clusters showing evidence of overconcentration are all lower-mass clusters.  For the four lowest mass clusters the average value of the concentration parameter $c_{200}$ is $11.6$, while for the six higher mass clusters the average value of $c_{200}$ is $4.4$.  $\Lambda$CDM would place $c_{200}$ between $3.4$ and $5.7$.
\end{abstract}


\section{Introduction}
	
	Galaxy clusters are the largest gravitationally-bound structures in the universe, and as such, can tell us many things about the origin and structure of the universe. Clusters indicate the locations of peaks in the matter density of the universe \citep{Allen11} and represent concentrations of dark matter.  Galaxy clusters are also places where gravitational lensing is likely to be observed \citep{Mollerach02,Kochanek03}. Gravitational lensing in a cluster can provide even more information, giving us a window not only to the cluster itself but to far more distant source galaxies.

	Gravitational lensing, both strong and weak, can be useful in the study of galaxy clusters.  Strong lensing, the formation of multiple resolved images of background objects by a cluster or other massive object, can be useful as it provides a direct measure of the mass contained within the Einstein radius of a cluster (e.g., Narayan \& Bartelmann 1997).  Weak lensing, the systematic but subtle change in ellipticities and apparent sizes of background galaxies, can also provide a precise measure of the mass of a cluster \citep{Mollerach02,Kochanek03}.  
	
		Galaxy cluster finding in optical data is performed using an algorithm based on some known properties of galaxy clusters \citep{Berlind06,Koester07a,Hao09,Soares10}.  \citet{Koester07a} describe the maxBCG method, a cluster-finding method used on the Sloan Digital Sky Survey (SDSS) data.  We describe in \S 3 how we used this method for cluster galaxy identification.
		
		The dark matter mass distribution of galaxy clusters is well fit by a Navarro-Frenk-White (NFW) profile \citep{NFW97,Wright00}.  One of the parameters in the NFW profile is the concentration parameter (here $c_{200}$), which is a measure of the halo density in the inner regions of the cluster.  The concentration parameter can be directly measured through weak lensing.  The standard cold dark matter cosmology ($\Lambda$CDM) describes the history of galaxy cluster formation and as such can make predictions (e.g., Duffy et al. 2008) about the value for $c_{200}$ as a function of cluster mass and redshift.  If measured values for $c_{200}$ are higher than predictions, then the clusters are said to be overconcentrated.  
			
	There have been indications from several groups \citep{Broad05,Broad08,Oguri09,Gralla11,Fedeli11,Oguri12} that galaxy clusters that exhibit strong lensing are overconcentrated. The overconcentration has been shown by disagreement between predicted and observed Einstein radii \citep{Gralla11} or by disagreement between predicted and observed concentration parameter \citep{Fedeli11,Oguri12}. There are some indications that the overconcentration problem is most significant in clusters with mass less than 10$^{14} h^{-1}M_{\sun}$ \citep{Fedeli11,Oguri12}.  This overconcentration problem might indicate that clusters are collapsing more than we expected \citep{Broad08}. This collapse may be related to baryon cooling, especially in the central galaxy of the cluster \citep{Oguri12}. 
		
	In this paper we describe a sample of ten galaxy clusters showing evidence of strong gravitational lensing.  These clusters were discovered in the SDSS during a search for strong lensing arcs.  We took follow-up data on these ten systems using the WIYN telescope at Kitt Peak National Observatory.  In this paper we describe our analyses of this data, including both the properties of the clusters and the properties of the arcs.  In \S 2 we address how these systems were found and how the data were taken.  We provide details regarding the searches which led to the discovery of these systems and we discuss the observing conditions at KPNO during the data acquisition.  In \S 3 we discuss identification of cluster members and measurements of cluster properties. We describe how we used the maxBCG method to identify cluster members, quantify cluster richness and estimate cluster masses.  We also describe how we found and used a scale factor to scale our richness measurements up to match those that would be measured in SDSS data.  We applied this scaling relation in order to use a relation between cluster richness and cluster mass that was calibrated using SDSS data.  Four of our ten systems are also included in \citet{Oguri12} and so we compare our results for cluster mass using cluster richness and strong lensing to their mass values which were found from strong and weak lensing.  In \S 4 we present measurements of the strong lenses.  For the lenses, we present Einstein radii, lens masses and lens velocity dispersions.  In \S 5 we discuss evidence of overconcentration.  We show that most of our clusters do not show evidence of overconcentration, but several of them do.  As the clusters showing evidence of overconcentration are all low mass clusters, they support recent results \citep{Oguri12,Fedeli11} suggesting that the overconcentration problem is most significant for lower mass clusters.     
	
	Throughout this paper, we assume a flat $\Lambda$CDM cosmology with $\Omega_m=0.3, \Omega_\Lambda=0.7$ and $H_{0}=100 \ h$ km  s$^{-1}$ Mpc$^{-1}$.

\section{Data Acquisition}
\subsection{Lens Searches}
	The Sloan Digital Sky Survey \citep{York00} is an ambitious endeavor to map more than 25$\%$ of the sky and to obtain spectra for more than one million objects. The SDSS was begun in 2000, and has completed phases I and II; phase III began in 2008 and will continue until 2014. The SDSS uses a $2.5$-m telescope located at Apache Point Observatory in New Mexico.  The Sloan Bright Arcs Survey (SBAS) is a survey conducted by a collaboration of scientists at Fermilab and has focused on the discovery of strong gravitational lensing systems in the SDSS imaging data and on subsequent analysis of these systems \citep{Allam07,Lin09,Diehl09,Kubo09,Kubo10,West12}.  To this point, the SBAS has discovered and spectroscopically verified 19 strong lensing systems with source galaxy redshift between $z=0.4-2.9$.  
	
\subsection{Follow-Up at WIYN:  Observing Details}	
	On February 26 and 27, 2009, we took follow-up data for ten of these systems at the 3.5-m Wisconsin-Indiana-Yale-NOAO (WIYN) telescope at Kitt Peak National Observatory.  The ten systems for which we took data for are listed in Table \ref{redshifts}.  We took follow-up data in order to obtain images with finer pixel scale, improved seeing, and fainter magnitude limits than were available in the SDSS data.  The pixel scale in the SDSS data is $0.396\arcsec$,  while the median seeing in the SDSS Data Release 7 (DR7) is $1.4\arcsec$ in the $r$ band.  Magnitude limits for DR7 are $22.2$, $22.2$, and $21.3$ in the $g$, $r$, and $i$ bands respectively.  
	
\begin{figure}
\begin{center}
\includegraphics[scale=0.8]{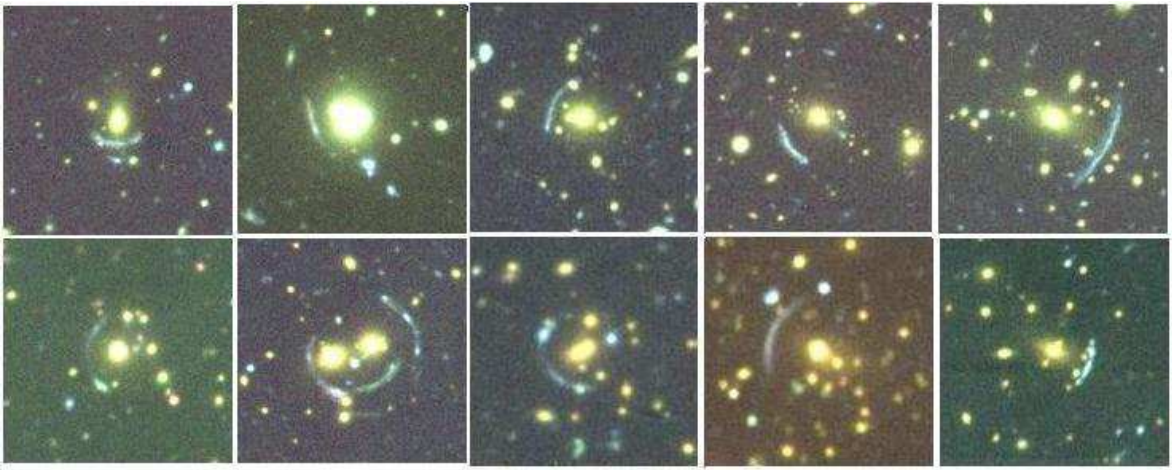}
\caption{A collage of the ten lensing systems.  The bottom row is (left to right):  SDSS J0901+1814, SDSS J1038+4849, SDSS J0900+2234, SDSS J1209+2640 and SDSS J0957+0509.  The top row is (left to right): SDSS J1511+4713, SDSS J1537+6556, SDSS J1439+3250, SDSS J1318+3942 and SDSS J1343+4155.   These images were produced by combining the stacked images in $g$, $r$ and $i$ filters.  Each image has dimensions of $49\arcsec \times 49\arcsec$.} \label{WIYNcollage}
\end{center}
\end{figure}

	The follow-up images were taken using the Mini-Mosaic camera, a camera that uses two CCDs, each of dimensions $2048\times4096$ pixels.  The pixel scale for the Mini-Mosaic camera is $0.14\arcsec$.  Images were taken using three filters, SDSS $g$, $r$ and $i$ filters.  A collage of color images of sections of these images showing the strong lenses is provided in Figure \ref{WIYNcollage}.
	
	For each data image, the exposure time was $450$-s and two exposures were taken for each field in each filter.  Later the exposures were stacked, leading to a total exposure time of $900$-s ($15$ minutes) per field in each filter.  Seeing was variable during the two nights, ranging from $0.49\arcsec$ for SDSS J1318+3942 to $1.54\arcsec$  for SDSS J1209+2640.  The median seeing was $0.74\arcsec$  for February 26 and $0.75\arcsec$  for February 27.  Magnitude limits for this data were estimated from the turnover in the number count histogram and were found to be approximately $24$ in $g$ band, $24$ in $r$ band and $23$ in $i$ band.   	
	
	The data were reduced using the NOAO Image Reduction and Analysis Facility \citep{Tody93}.  The data were flat-fielded using both dome flats taken on site and superflats produced from data images.  Cosmic rays were removed using the IRAF task LACosmic \citep{vanDok01}.  IRAF was again used to make corrections to the world coordinate system and to stack the two images in each filter.  Object magnitudes were measured in several measurement apertures using SExtractor \citep{Bertin96}.  Finally, the instrumental magnitudes measured by SExtractor were converted to calibrated magnitudes.  This was done by finding the model magnitudes of stars in the SDSS DR7 Catalog Archive Server that also appeared in the WIYN data and finding the offset in magnitudes in the $g$, $r$, and $i$ bands.  The median offset in each filter for each field was then added to the SExtractor magnitudes (using $\tt{MAG\_AUTO}$).  
	
\section{Galaxy Cluster Properties}
\subsection{Identifying Cluster Galaxies}  
		We first sought to characterize richness of the clusters in terms of $N_{gals}$, the number of cluster members within 1 $h^{-1}$ Mpc of the brightest cluster galaxy (BCG) \citep{Hansen05} by using the maxBCG method.  The maxBCG method \citep{Koester07a} uses three primary features of galaxy clusters to facilitate the detection of clusters in survey data.  First, galaxies in a cluster tend to be close together near the center and to become more separated from one another toward the outskirts of the cluster.  Second, galaxies in a cluster tend to closely follow a sequence in a color-magnitude diagram; this is referred to as the E/S0 ridgeline, where E and S0 refer to galaxy types in the Hubble classification.  Finally, galaxy clusters typically contain a central BCG, which is defined as the brightest galaxy in the cluster.  In all of the clusters in our sample, one or two BCGs can be seen near the center of the cluster surrounded by lensing arcs.  While the dark matter halo dominates the lensing potential, the BCG contributes to the lensing potential as well since it comprises a large fraction of the baryonic matter in the cluster.  Typically the BCG would be expected to have a color similar to that of the other cluster galaxies and to be almost at rest with respect to the halo of the cluster.
	
	Considering these properties of cluster galaxies, we searched the SExtractor catalog files for objects that: (1) were classified as galaxies, not stars, (2) were within 1 $h^{-1}$ Mpc of the central BCG, (3) had the characteristic $r-i$ and $g-r$ color of the E/S0 ridgeline, and (4) met a particular magnitude limit.

\begin{figure}
\begin{center}
\includegraphics[scale=0.8, angle=90]{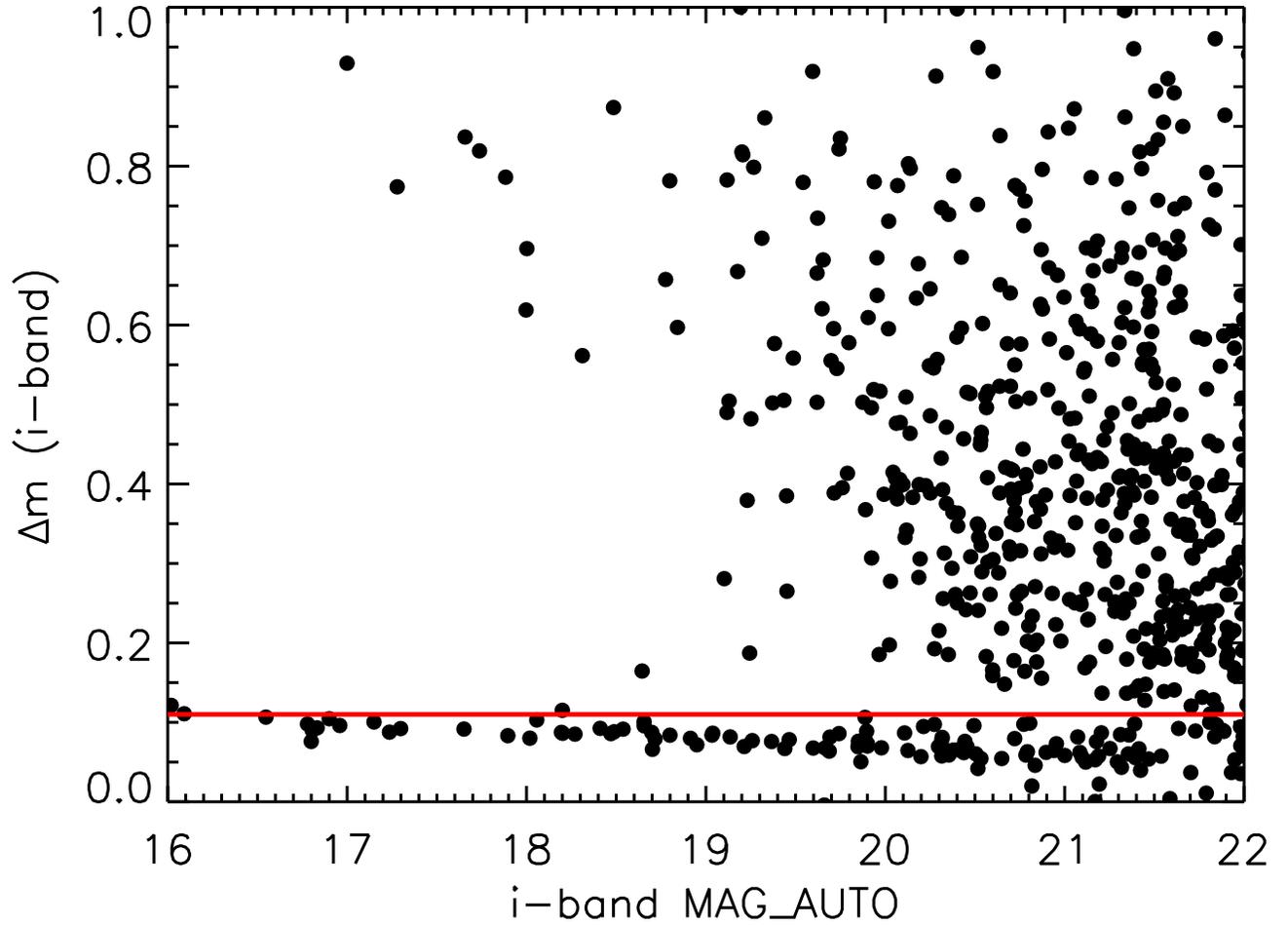}
\caption{A sample plot of $\Delta$m vs. i-band $\tt{MAG\_AUTO}$ for SDSS J1439+3250.  Recall that $\Delta$m is 2$\arcsec \ \tt{MAG\_APER}-\tt{MAG\_AUTO}$.  The horizontal red line is the star-galaxy cutoff we used, meaning objects with $\Delta$m$\leq$0.11 were cut as stars.}\label{star}
\end{center}
\end{figure}

	In order to separate galaxies from stars, we compared two different SExtractor magnitudes, $\tt{MAG\_AUTO}$ and $\tt{MAG\_APER}$.  $\tt{MAG\_AUTO}$ is the flux measured above background in a variable-size elliptical aperture.  $\tt{MAG\_APER}$ uses a circular aperture of fixed size to determine magnitude; we used a diameter of $2.0\arcsec$.  The difference $\tt{MAG\_APER} - \tt{MAG\_AUTO}$ (henceforth $\Delta m$), can be used to identify the galaxies: stars stand out from galaxies because stars typically have a nearly identical shape while galaxies generally do not.  Thus for stars the fixed aperture of $\tt{MAG\_APER}$ will measure a fairly constant fraction of the light that the variable aperture of $\tt{MAG\_AUTO}$ will measure.  Therefore, the difference between the measurements ($\Delta m$) will be mostly constant for stars, but not for galaxies.  We used this fact to find stars by plotting $\Delta m$ vs.  $\tt{MAG\_AUTO}$.  In this plot, stars will be found on a mostly horizontal line of nearly constant $\Delta m$ value; this line is referred to as the stellar locus (see Figure \ref{star}). 
	
	We also tried using the SExtractor parameter $\tt{CLASS\_STAR}$ for star-galaxy separation by requiring $0 \leq \tt{CLASS\_STAR} \leq 0.9$ (1 is highly star-like and 0 is highly galaxy-like in this parameter) and remeasuring $N_{gals}$ with this requirement.  We chose this cutoff because when we plotted $\tt{CLASS\_STAR}$ against i-band magnitude ($\tt{MAG\_AUTO}$), we found a tight stellar sequence within $0.1$ of $\tt{CLASS\_STAR}=1$.  We found that the mean difference in $N_{gals}$ values was $0.3$, which corresponds to a mean percent difference of $1.7\%$.  Thus we conclude that the $\Delta m$ cut method is equivalent to using $\tt{CLASS\_STAR}$.

\begin{figure}
\begin{center}
\includegraphics[scale=0.7, angle=90]{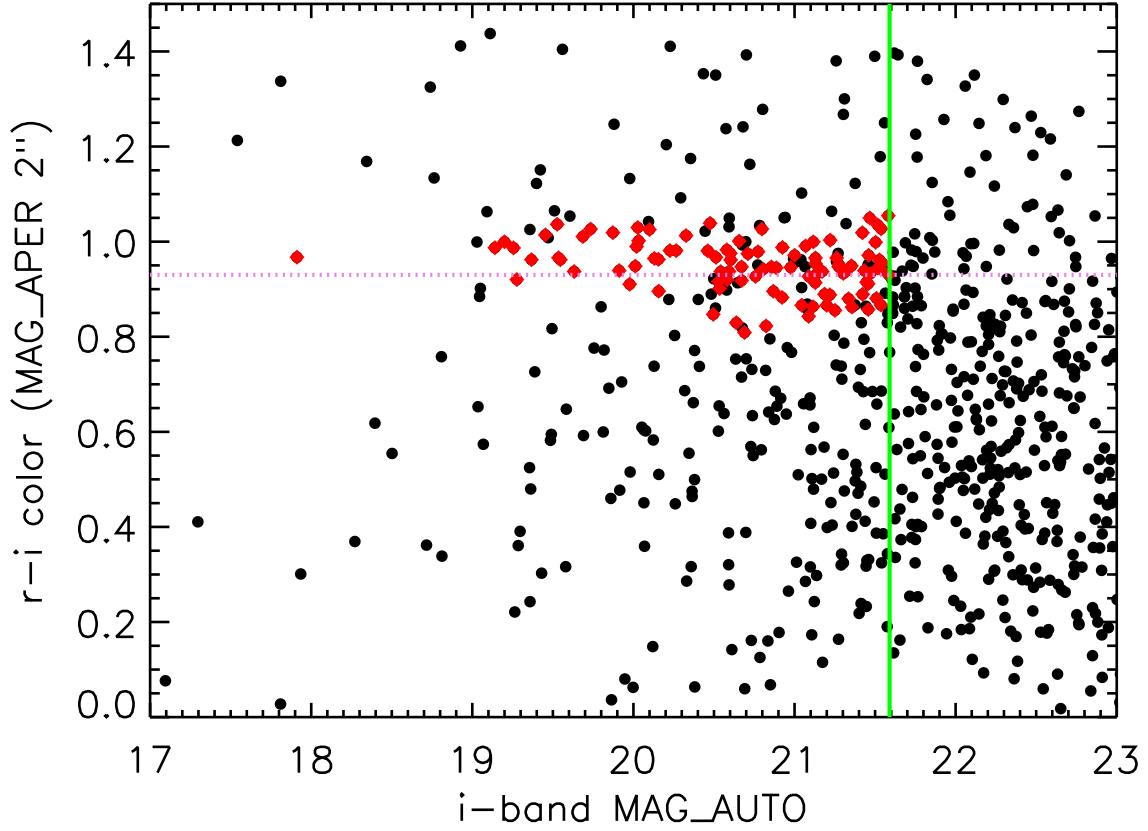}
\caption{The $r-i$ color-magnitude diagram for SDSS J1209+2640.  The black dots denote the galaxies, the red diamonds denote the cluster galaxies, the vertical green line shows the value of $0.4L^*$ and the horizontal violet dotted line represents the red sequence $r-i$ color.  The objects plotted are galaxies within 1 $h^{-1}$ Mpc of the BCG.}\label{colormag}
\end{center}
\end{figure}

		In order to select galaxies that are members of the cluster, we used the red sequence method \citep{Gladders00,Koester07a}.  This approach involves plotting a color-magnitude diagram of the $g-r$ and $r-i$ colors of the galaxies vs. their $i$-band magnitude, looking for a nearly horizontal line of galaxies of similar color. Galaxies in a cluster are at similar redshifts and will be largely coeval, leading them to have similar colors.  Thus the galaxies that populate the red sequence are likely to be cluster members.  For each cluster we identified the $g-r$ and $r-i$ color of the red sequence on the plots.  A sample color-magnitude diagram is shown in Figure \ref{colormag}.
	
	We also used a second method to check our identification of the red sequence color.  For both $g-r$ and $r-i$ colors, we made a histogram of the colors of the galaxies within 1 $h^{-1}$ Mpc of the BCG and found the distribution near the red sequence color we had previously identified.  We then fit this section of the histogram with a Gaussian profile and found the mean color of the red sequence galaxies.       

	Ultimately we used the first method (color-magnitude diagrams) to obtain a reasonable range of values for the colors of the red sequence and we used the second method (histograms) to determine final values for the colors.  When we made color cuts, we only allowed galaxies that were within 2$\sigma$ of the $r-i$ and $g-r$ colors, where $\sigma$ was defined as:
\begin{equation}
\sigma = \sqrt{\begin{array}{c}{(\sigma_{intrinsic})^2}+(\sigma_{color})^2\end{array}}   \label{errcolor}
\end{equation}      
Here $\sigma_{intrinsic}$ is the intrinsic scatter in the red sequence color in the absence of measurement errors, which we took to be $0.06$ for $r-i$ and $0.05$ for $g-r$ \citep{Koester07a}. $\sigma_{color}$ is the color measurement error found by adding the SExtractor aperture magnitude measurement errors in quadrature.  

	Finally we cut any galaxies that had a magnitude dimmer than $0.4L^*$, where  $L^*$ is defined as the luminosity at which the luminosity function \citep{Schechter85} changes from a power law to an exponential relation.  In the maxBCG algorithm $0.4L^*$ is used as a limiting magnitude \citep{Koester07b}, and so we adopt this as our magnitude limit as well.  We referred to a table of $0.4L^*$ \citep{Annis10} as a function of $z$ to make cuts, allowing only galaxies brighter than $0.4L^*$ in $i$-band.  All values used for cluster galaxy cuts are provided in Table \ref{colors}.   
	
\subsection{Cluster Properties}
\subsubsection{Area Corrections}
\begin{figure}
\begin{center}
\includegraphics[scale=0.8]{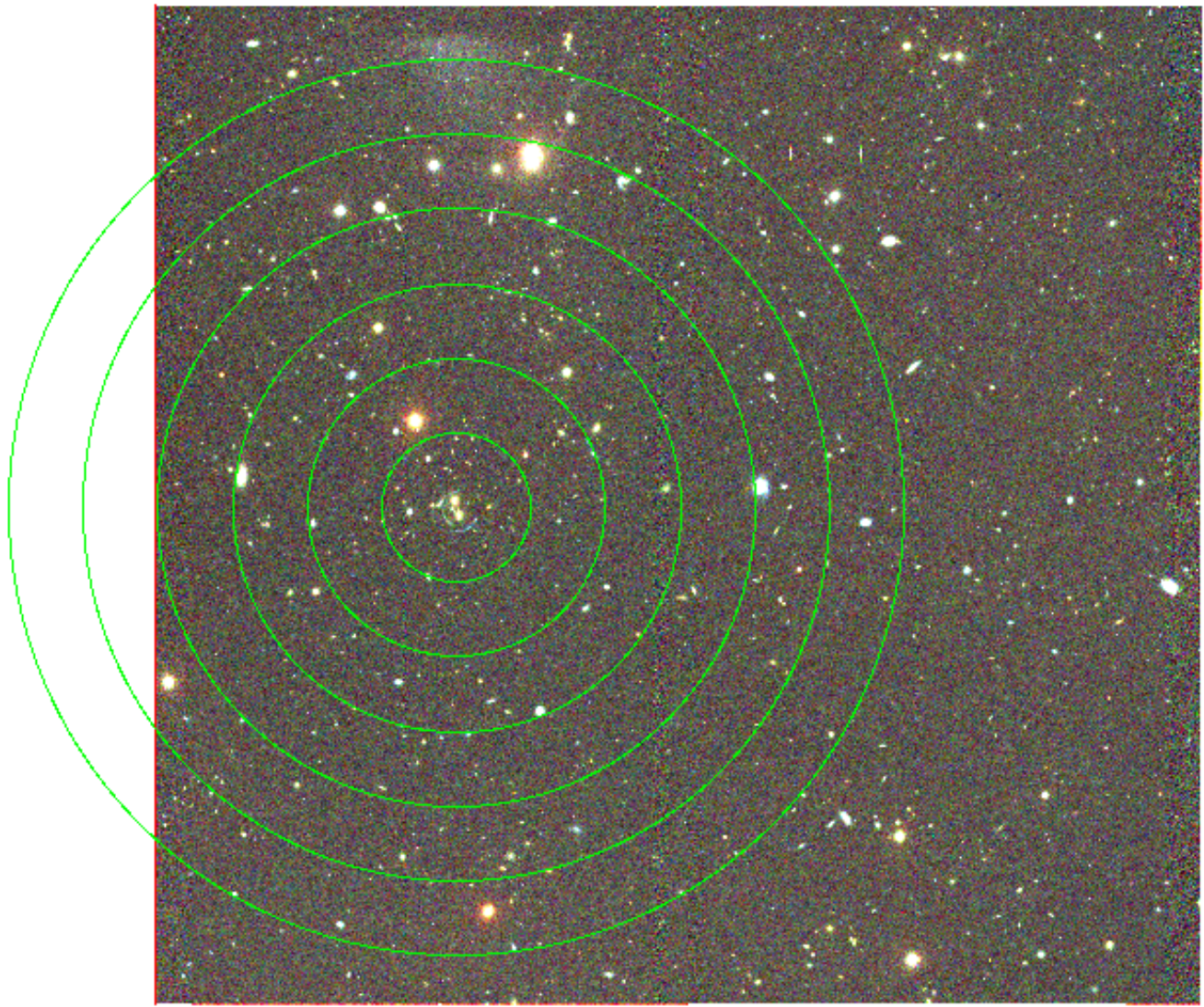}
\caption{An image of SDSS J1038+4849, with a circular region of radius $1 \ h^{-1}$ Mpc centered on the BCGs.  We have divided the aperture into six annuli in order to apply Equation \ref{extrapolate} for area corrections.}\label{annuli}
\end{center}
\end{figure}

  We applied the four cuts described in $\S 3.1$ to measure $N_{gals}$.  However we found that for several of the ten systems, regions of the cluster were not in the image.  The reason for this is that when we took the data, our primary focus was on the strong lensing arcs, which were near the center in all of our images. In order to address this problem and still obtain accurate values for $N_{gals}$, we extrapolated values for $N_{gals}$ in the area off the CCD. In order to do this, we divided the 1 $h^{-1}$ Mpc aperture into six annuli with constantly increasing radii, as shown in Figure \ref{annuli}. We assumed that the number of galaxies in each annulus should only be a function of radius; this would suggest that the number of galaxies per area should be a constant in each annulus.  Mathematically,
\begin{equation}
N_{total}=N_{on\_CCD}\left(
\frac{A_{ann}}{A_{ann\_on\_CCD}}\right) \label{extrapolate}
\end{equation}			
where $N_{total}$ means the total number of galaxies in each annulus, $N_{on\_CCD}$ means the number of galaxies actually found in the image in each annulus, $A_{ann}$ means the area of the annulus and $A_{ann\_on\_CCD}$ means the area of the annulus that was on the CCD.  

\begin{figure}
\begin{center}
\includegraphics[scale=0.7, angle=90]{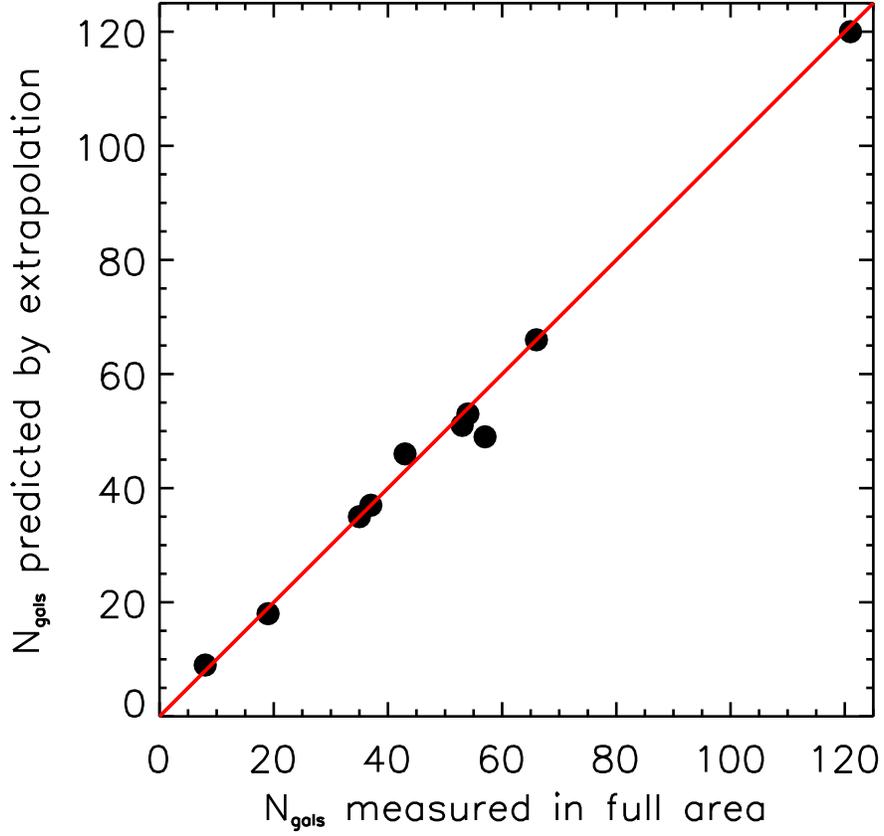}
\caption{This plot is a test of the accuracy of the $N_{gals}$ extrapolation described in Equation \ref{extrapolate}.  Here we plot $N_{gals}$ values measured in SDSS data, with the measured values on the x-axis and the predicted values on the y-axis.  The red line is the $y=x$ line.  Since the data closely follow the $y=x$ line, we conclude that the predictions from the extrapolation are quite accurate.}\label{sloan}
\end{center}
\end{figure}

		We checked the accuracy of Equation \ref{extrapolate} using the SDSS data.  We measured $N_{gals}$ twice, once covering the full 1 $h^{-1}$ Mpc (taking this as true $N_{gals}$) and once covering only as much of the 1 $h^{-1}$ Mpc as was on the CCD in the WIYN data.  We then used Equation \ref{extrapolate} to predict the final values of $N_{gals}$ based on the measurements with the WIYN area cuts.  Finally we compared the predicted values for $N_{gals}$ to the measured (true) values and found them to be similar.  We plot the two sets of $N_{gals}$ against each other in Figure \ref{sloan}.  Note that the points follow the $y=x$ line very closely, indicating that the measured and extrapolated values are quite similar and suggesting that the richness extrapolation works well.  The typical fractional error in the extrapolated values is $0.06$. 

\subsubsection{Richness Measurements} \label{N200}	
	We next found the richness, $N_{200}$ \citep{Hansen05}, the number of galaxies in a spherical region within which the density was 200$\rho_{crit}$, where $\rho_{crit}$ is the critical density of the universe.  The radius of this spherical region of space is termed $r_{200}$.  \citet{Hansen05} give $r_{200}$ as:
\begin{equation}
r_{200}=0.156(N_{gals})^{0.6} h^{-1}Mpc
\end{equation}	 
We used the area-corrected values for $N_{gals}$ when calculating $r_{200}$.  In order to find $N_{200}$ we again applied the four cuts discussed in $\S3.1$, this time using $r_{200}$ as the distance cut rather than 1 $h^{-1}$ Mpc.  Finally, once we found $N_{200}$, we again applied the area corrections using Equation \ref{extrapolate}. 
	
	We used the variable elliptical aperture of $\tt{MAG\_AUTO}$ and the circular $2\arcsec$ and $3\arcsec$ diameter apertures using $\tt{MAG\_APER}$ in order to determine object magnitudes and thus colors.  We used $2\arcsec$ and $3\arcsec$ because both were significantly larger than the seeing FWHM, for which the median value was about $0.75\arcsec$.  The differences in colors measured in different apertures were usually small, on the order of $0.05$ magnitudes, but could be up to $0.2$ magnitudes.  Since identification of a cluster galaxy depends on color, there was a resulting variation in richness values for different apertures.  We determined that the $2\arcsec$ aperture had the highest signal to noise by comparing the measurement errors of the $g-r$ and $r-i$ colors to see in which aperture the errors were typically lowest.  We found that the 2$\arcsec$ aperture typically had the lowest error value; therefore we used the colors and thus richness values in the 2$\arcsec$ aperture for richness measurements.  However, we considered the variation in richness values to determine the error in richness:  we took the standard deviation of the three values for $N_{200}$ for each cluster and used these values for the uncertainty in $N_{200}$.  
	
\subsubsection{Cluster Mass}
	We define $M_{200}$ to be the mass contained within a spherical region of radius $r_{200}$ \citep{Johnston07}.  An empirical relation between mass and richness is found in \citet{Johnston07} using a large sample of maxBCG clusters from the SDSS:
\begin{equation}
M_{200}(N_{200})=M_{200|20}\left(\frac{N_{200}}{20}\right)^{\alpha_N}  \label{Johnston}
\end{equation}	 
In this equation $M_{200|20}=(8.8 \pm 0.4_{stat}\pm 1.1_{sys})\times10^{13}h^{-1}M_{\sun}$ and $\alpha_{N}=1.28 \pm 0.04$.  Equation \ref{Johnston} was found empirically using data from the SDSS, using mean redshift of $z=0.25$.  

	The error in $M_{200}$ values was considered in \citet{Rozo09}.  In that paper, the logarithmic scatter in mass at fixed richness is given as:
\begin{equation}
\sigma_{lnM|N}=0.45_{-0.18}^{+0.20}
\end{equation}	  
We thus can approximate the uncertainty in the mass itself as:
\begin{equation}
\Delta M=0.45M_{200} \label{Rozo}
\end{equation}	  
We also propagate error from the uncertainty in values of $N_{200}$ through equation \ref{Johnston}.  Our final values for error on $M_{200}$ were found by adding the uncertainty in the mass and the propagated error in quadrature.  The propagated fractional errors had a median value of $0.13$ while the scatter described by Equation \ref{Rozo} had a value of $0.45$.  The combined fractional errors had a median value of $0.47$, with the scatter in mass dominating the errors. 

\subsubsection{Velocity Dispersion}
	\citet{Becker07} give an empirical relationship for velocity dispersion as a function of richness found from redshifts of cluster members in the maxBCG cluster sample:
\begin{equation}
\langle\ln\sigma_v\rangle=A+B\ln\frac{N_{200}}{25} \label{dispersion}
\end{equation}	  
The constants $A$ and $B$ are referred to as mean-normalization and mean-slope, respectively.  They are given as $A=6.17 \pm 0.04$ and $B=0.436 \pm 0.015$.  \citet{Becker07} also found a relation for the scatter, $S$, in the velocity dispersion.  The scatter is defined to be the standard deviation in $\ln{\sigma_{v}}$:
\begin{equation}
S^2=C+D\ln\frac{N_{200}}{25}   \label{disp_err}
\end{equation}	 
where $C=0.096 \pm 0.014$ and $D=-0.0241 \pm 0.0050$.  We used this relation to calculate the errors on the velocity dispersion values, defining the errors as one standard deviation.  We also propagated the error on $N_{200}$ through Equation \ref{dispersion} and added these errors in quadrature to the errors found from Equation \ref{disp_err}.  Again the propagated errors are minimal:  The median fractional error on the velocity dispersions from the propagated error on $N_{200}$ is $0.08$, while the median fractional error from Equation \ref{disp_err} is $0.31$, leading to an overall median fractional error of $0.33$.  
	     	
\subsubsection{Errors on Richness and Mass}  \label{Erronrichmass}

\begin{figure}
\begin{center}
\includegraphics[scale=0.9, angle=90]{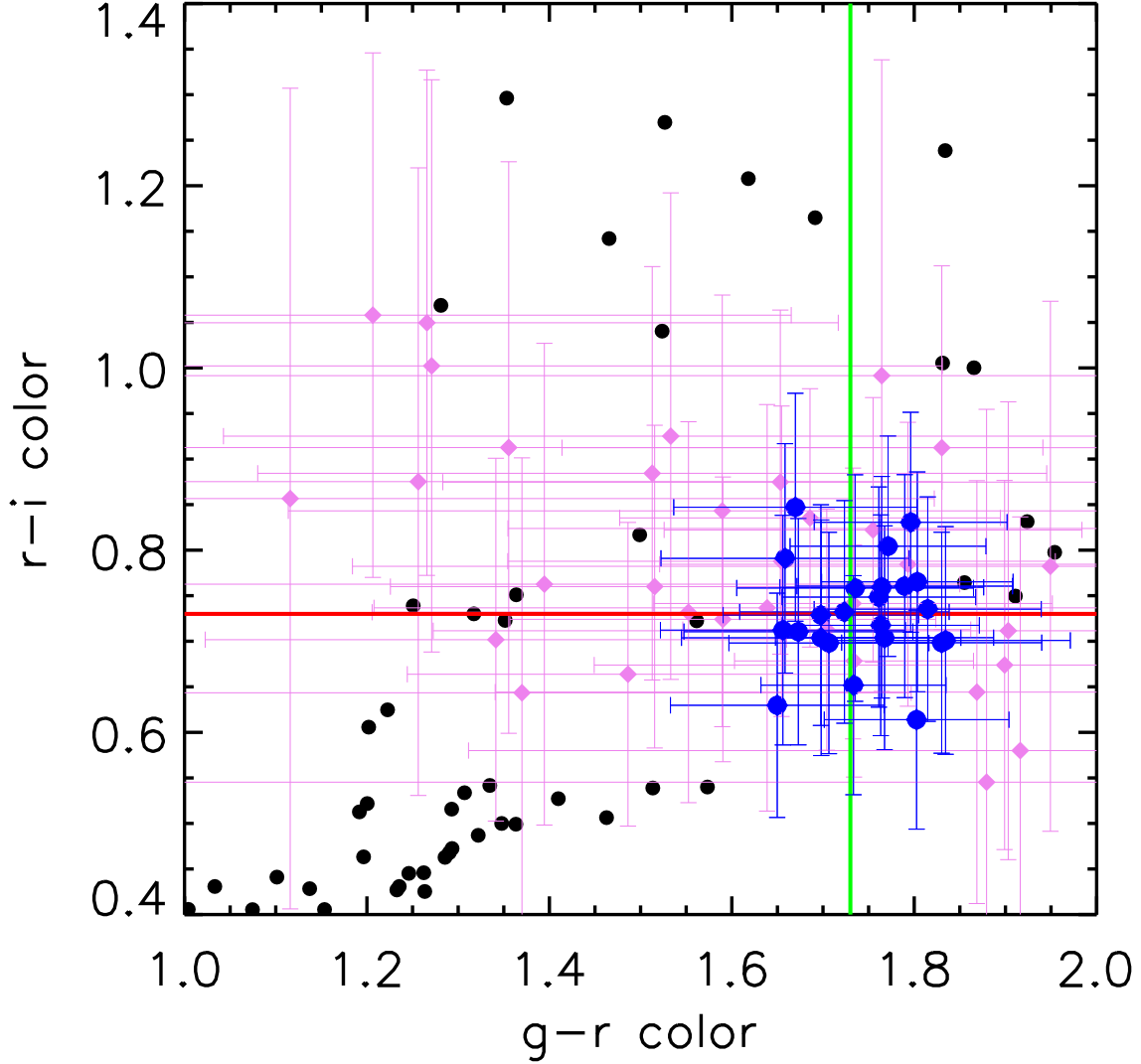}
\caption{A comparison of cluster members found in our data (2$\arcsec \ \tt{MAG\_APER}$) and in the SDSS data for SDSS J1318+3942.  The larger blue circles represent cluster members found in our data and the smaller violet diamonds are cluster members found in the SDSS data.  The smallest black circles are all galaxies within 1 h$^{-1}$ Mpc of the BCG that are brighter than $0.4L^*$ in our data but do not meet the color cuts to be considered cluster members.  The error bars represent $2\sigma$, with $\sigma$ defined by Eq. \ref{errcolor}.  The solid lines mark the cluster red sequence colors for the WIYN data:  the vertical green line marks the g-r color and the horizontal red line marks the r-i color.}\label{clustermem}
\end{center}
\end{figure}

	In order to better constrain the error on our richness measurements, we also measured colors and richnesses for the 10 systems using the SDSS data.  We found that richness values from the SDSS are typically much higher than those found in this paper; the mean ratio of $N_{gals}$(SDSS) to $N_{gals}(2\arcsec \ \tt{MAG\_APER}$) is $1.75$ (for WIYN $N_{gals}$ before area corrections, using only cluster area found both in WIYN and SDSS data; see Table \ref{richvals}).  These differences apparently arise because there is a larger error in magnitudes measured in the SDSS than in the data used here.  This allows some objects to be counted as cluster members in the SDSS that are not counted as cluster members in the WIYN data.  Note that in Figure \ref{clustermem}, a color-color diagram for SDSS J1318+3942, more cluster members are found in SDSS data, but those objects are much more scattered in color-color space and many are not true cluster members.  On the other hand, fewer objects are found in the WIYN data, but these objects form a much tighter red sequence and are more likely to be genuine cluster members.  
	
\begin{figure}
  \begin{minipage}[t]{3in} 
    \centering
    \includegraphics[width=2.6in,angle=90]{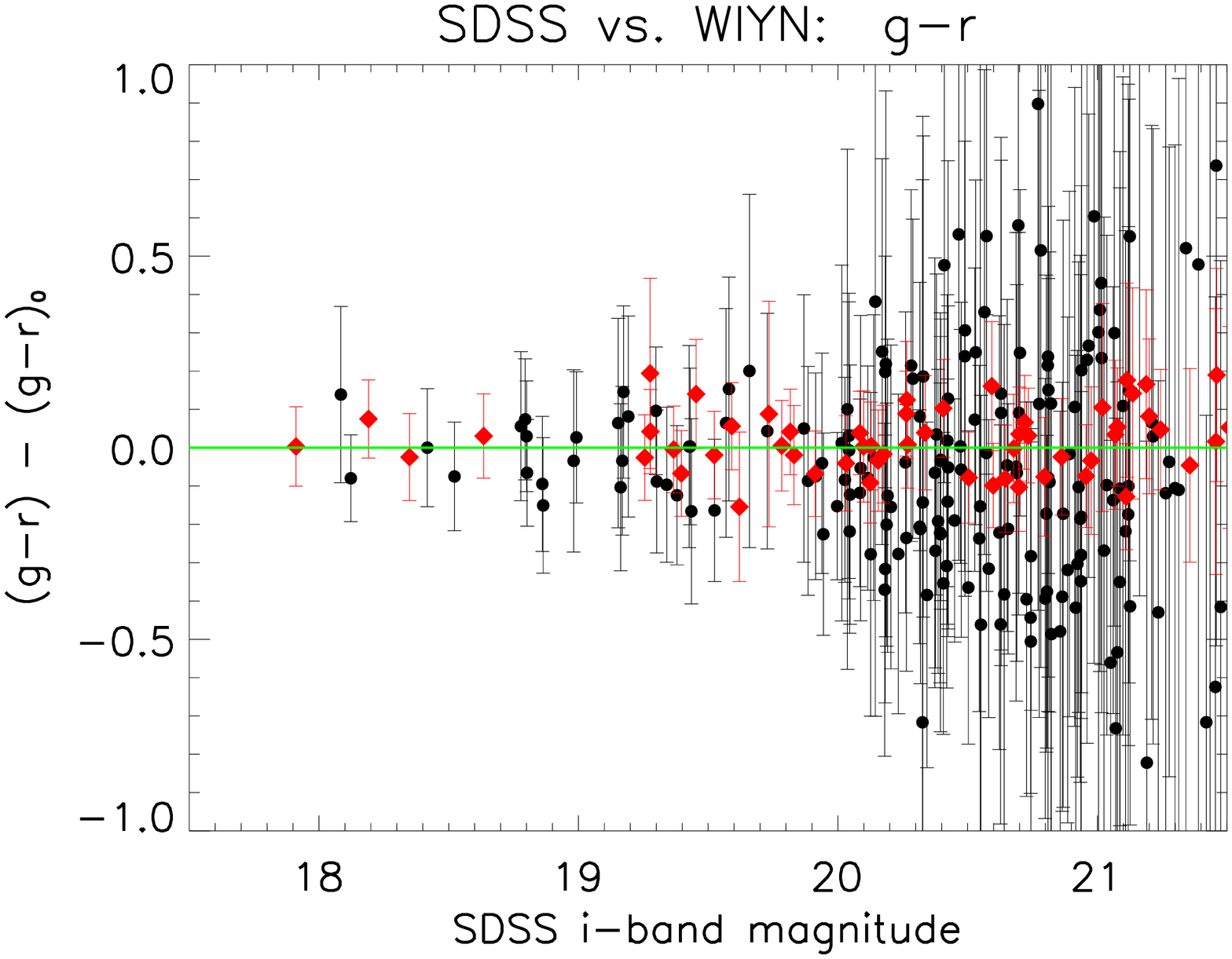}
     \label{color_gr}
  \end{minipage}
  \hspace{0.3in}
  \begin{minipage}[t]{3in} 
    \centering
    \includegraphics[width=2.5in,angle=90]{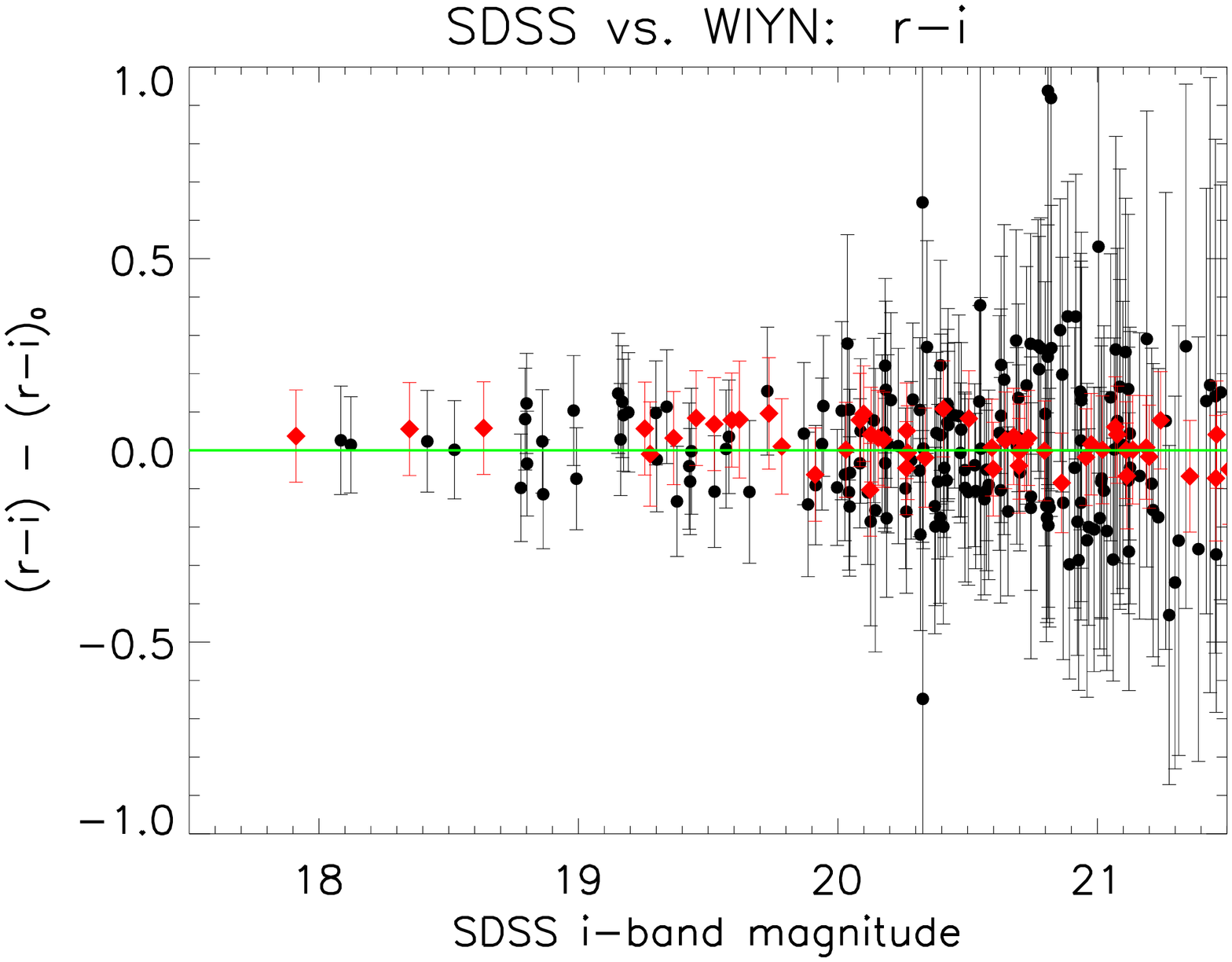}
  \end{minipage}
  \caption{A plot of color difference versus SDSS i-band magnitude for all cluster members in both SDSS and WIYN data.  Color difference is defined as the difference between the actual $r-i$ or $g-r$ color of each cluster galaxy and the measured red sequence color for that cluster.  Cluster galaxies in each of the ten clusters are plotted together here.  The red diamonds denote WIYN data points and the black circles denote SDSS data points.  The error bars represent $2\sigma$, where $\sigma$ is defined by Eq. \ref{errcolor}.  For SDSS measurements, color is found from SDSS model magnitudes and the red sequence colors were measured in SDSS data.  For WIYN measurements, color is found from $2\arcsec$ $\tt{MAG\_APER}$ magnitudes and red sequence colors were measured in WIYN data.  Note that WIYN data points are found much closer to the central line that represents color difference of $0$, while SDSS points can be found further away.  To be counted as cluster members, points must be within $2\sigma$ of the cluster red sequence colors, but $2\sigma$ is larger for the SDSS points.}
\end{figure}

	We also include Figure \ref{color_gr}, in which we show the deviation of each cluster galaxy's color from the measured color of the E/S0 ridgeline; we plot this vs. SDSS i-band magnitude for all ten clusters.  We found g-r and r-i colors for objects considered to be cluster galaxies within 1 $h^{-1}$ Mpc of the BCG in WIYN data and in SDSS data and compared them to the characteristic red sequence colors of the respective clusters.  We also found the errors in colors for both sets of data using Equation \ref{errcolor} to find $\sigma$.  We used magnitude errors reported by SExtractor for WIYN data and errors on model magnitudes for SDSS data.  The error bars shown represent $2\sigma$.  It can be seen in Figure \ref{color_gr} that the differences between the measured color and the cluster color are much larger in the SDSS data than in WIYN data but the errors are larger for SDSS data as well.  Due to these larger errors in SDSS data, there is a higher likelihood that objects with larger color deviations will still be counted as cluster members.  
	
	The differences in richness values between WIYN and SDSS data persist even at bright magnitudes.  We measured values for $N_{gals}$ at an i-band magnitude of $19.38$, which is the value for $0.4L*$ corresponding to $z=0.25$. We found that the mean ratio of $N_{gals}$(SDSS) to $N_{gals}(2\arcsec \ \tt{MAG\_APER}$) is $1.63$, meaning that SDSS values are typically about $60\%$ higher than WIYN values.  Thus we find that in general for these ten clusters richness values measured in our data do not closely match values measured in the SDSS data.
	
\begin{figure}
\begin{center}
\includegraphics[scale=0.7, angle=90]{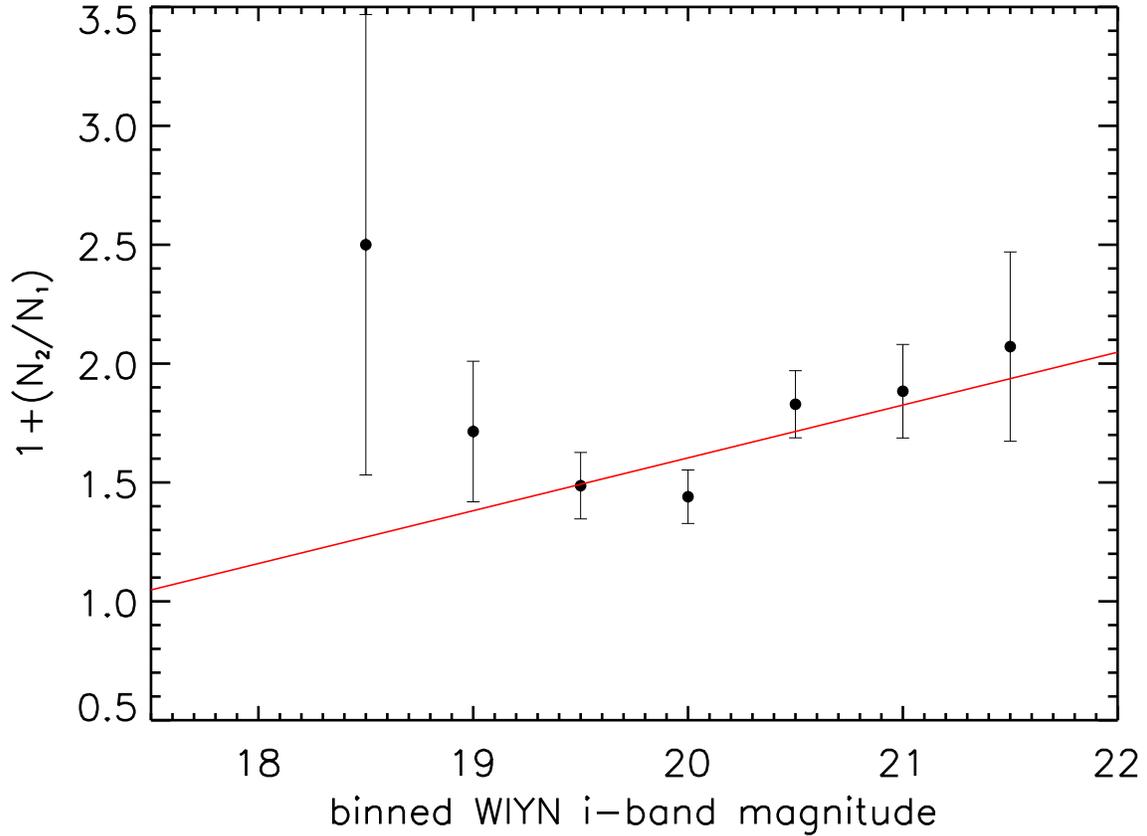}
\caption{A plot comparing objects counted as cluster galaxies only in SDSS data and in both WIYN and SDSS data.  Here $N_1$ is the number of cluster members found in both SDSS and WIYN in that magnitude bin and $N_2$ is the number of cluster members found only in SDSS.  We plot the ratio $1+\frac{N_2}{N_1}$ (which we refer to in the text as $C$) on the y-axis and the magnitude bin on the x-axis, where magnitude bins are $0.5$ magnitude in size.  The red line is a linear best fit, found using IDL routine $\tt{FITEXY}$.  The equation of that line is $C=(0.222 \pm 0.116)m_{i\_WIYN}+(-2.84 \pm 2.29)$, where $m_{i\_WIYN}$ represents magnitude in $i$-band $\tt{MAG\_AUTO}$.}  \label{failure}
\end{center}
\end{figure}

	However, since the mass-richness relation (Equation \ref{Johnston}) is calibrated from SDSS data, if we use WIYN richness values with this equation, we would expect the masses to be biased to be too low.  Therefore, we determined it would be necessary to scale our measured richness values up to match SDSS values.  To do that, we we first found all objects that were counted as cluster galaxies ($N_{gals}$) only in WIYN (not in SDSS) and then found the opposite, objects counted as cluster galaxies only in SDSS but not in WIYN.  We then also found the galaxies counted as cluster galaxies in both WIYN and SDSS.  Our goal was to constrain the amount that SDSS was overcounting galaxies.  To do that we found the ratio
\begin{equation}
C=\frac{N_1+N_2}{N_1}=1+\frac{N_2}{N_1}  								\label{ratios}
\end{equation}
where $N_1$ represents the number of cluster members found in both WIYN data and SDSS data and $N_2$ represents the number of cluster members found only in SDSS data.  Since we expect the numbers of galaxies in each magnitude bin to be a Poisson distribution, the standard deviation on $N_1$ and $N_2$ would be simply the square root of each.  Then the fractional error on Equation \ref{ratios} would be
\begin{equation}
\sigma_{C} = \frac{N_2}{N_1}\sqrt{\frac{1}{N_2}+\frac{1}{N_1}}
\end{equation}
We then plotted $C$ against binned WIYN i-band ($\tt{MAG\_AUTO}$) model magnitude.  The result is shown in Figure \ref{failure}.  We fit the data with a linear relation using IDL routine $\tt{FITEXY}$, which applies a linear fit including error bars.  The final relation found was
\begin{equation}
C=(0.222 \pm 0.116)m_{i\_WIYN}+(-2.84 \pm 2.29)  \label{scaler}
\end{equation}
The magnitude $m_{i\_WIYN}$ is WIYN $i$-band magnitude from $\tt{MAG\_AUTO}$.  When this equation is evaluated at i-band $m=19.38$, the value for $0.4L*$ at the mean SDSS redshift of $0.25$, then $C=1.47$.  We took this as the correction factor for our richness values.  

	We measured $N_{gals}$ and corrected these values for missing area in WIYN using Equation \ref{extrapolate}.  Then we included the above correction factor when calculating $r_{200}$, letting
\begin{equation}
r_{200}=0.156(N_{gals\_SDSS})^{0.6}=0.156(1.47N_{gals\_WIYN})^{0.6}  
\end{equation}
We remeasured $N_{200}$ using the new value for $r_{200}$ and corrected for missing area.  Finally we scaled these new $N_{200}$ values by multiplying them by the same scale factor of $1.47$.  We used these scaled values of $N_{200}$ to find $M_{200}$, velocity dispersion and concentration parameter.  We give values for all quantities found without the scale factor in Table \ref{clusters} and we give the values found with the scale factor in Table \ref{scaled}.  

	We find the scaled values for $N_{200}$ are on average $1.7$ times bigger than the unscaled values.  This leads the new values for $M_{200}$ (those found from the scaled richness values) to be $2.0$ times larger than the previous values.  Also new values for velocity dispersion are $1.3$ times larger than previous values, while new values for concentration parameter are all smaller, on average $0.63$ times the previous values (see $\S$ \ref{final}).   

\subsubsection{Comparison of Results}
	Several other groups have measured cluster masses or related quantities for some of our clusters.  \citet{Oguri12} present combined strong and weak lensing analyses for $28$ clusters, including $4$ of the clusters discussed in this paper.  This allowed us to compare our results for $M_{200}$ to their results for these four systems.  As \citet{Oguri12} present values for $M_{vir}$, we converted these to $M_{200}$ values using the method described in Appendix A of \citet{Johnston07} (see \S \ref{overconc}).   
	
\begin{figure}
\begin{center}
\includegraphics[scale=0.8, angle=90]{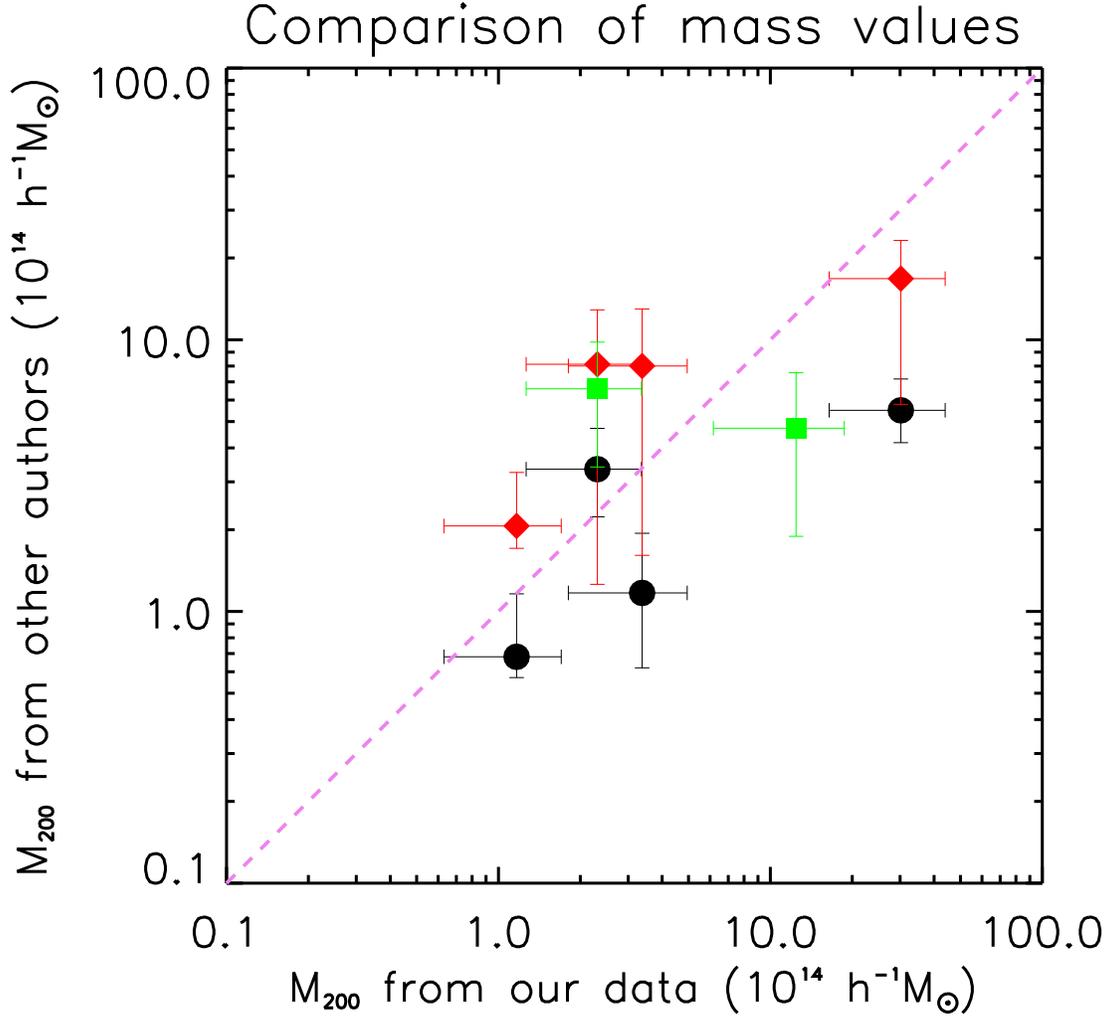}
\caption{A comparison of $M_{200}$ values in this paper and in other papers.  The black circles represent mass values in \citet{Oguri12}, the red diamonds represent mass values in \citet{Bayliss11} and the green squares represent mass values in \citet{Drabek12}.  The dotted violet line is the $y=x$ line.}\label{mass}
\end{center}
\end{figure}

	\citet{Bayliss11} provided velocity dispersions for $4$ of our clusters.  We used the relation between cluster mass and galaxy velocity dispersion given in \citet{Evrard08} to find $M_{200}$:
\begin{equation}
b_v^{\frac{1}{\alpha}}M_{200c}=10^{15}M_\sun\frac{1}{h(z)}\left(\frac{\sigma_{gal}}{\sigma_{15}}\right)^{\frac{1}{\alpha}}
\end{equation}	
Here $h(z)$ is the Hubble parameter, $b_v = \sigma_{gal} / \sigma_{DM}$ is the velocity bias (we assume $b_v = 1$), $\sigma_{gal}$ is the galaxy velocity dispersion, $\sigma_{DM}$ is the dark matter velocity dispersion,  $\sigma_{15} = 1084 \pm 13$~km~s$^{-1}$, and $\alpha = 0.3359 \pm 0.0045$.  \citet{Drabek12} present masses for two clusters, SDSS J1343+4155 and SDSS J1439+3250, based on spectroscopy of a sample of galaxies in these clusters.  We summarize all the values of $M_{200}$ found by these groups in Table \ref{masscomp}.  In Figure \ref{mass}, we plot the $M_{200}$ values from the three other papers against our $M_{200}$ values; the dotted line in the plot is the $y=x$ line.  We find that our values are reasonable in light of the findings of other groups as when we plot our values against those from other groups, the points are all scattered around the $y=x$ line.  

\section{Strong Lensing Properties}
	In a strong lensing system, if the source galaxy and the galaxy cluster are perfectly aligned, then the image formed will be a perfect ring, or Einstein ring.  The radius of this ring is referred to as the Einstein radius.  The Einstein radius for a symmetric mass distribution treated as a thin sheet is given by \citep{Narayan97}:
\begin{equation}
\theta_E=\sqrt{{\frac{4GM}{{c^2}}}{\frac{D_{ds}}{D_{d}D_{s}}}} \label{eqn_eins}
\end{equation}	 	
where $D_{d}, D_{s}$, and D$_{ds}$ are angular diameter distances to the lens, to the source, and from lens to source, respectively, $c$ is the speed of light, $G$ is the gravitational constant, and $M$ is the mass contained within the Einstein radius.  We measured the Einstein radius of each of the clusters directly by fitting a circle to the visible arc and measuring the radius of that circle.  We intend in the near future to apply more sophisticated mass models to the arcs in order to better characterize the Einstein radii, but this method provides an estimate.  The values found here are all very similar to those presented in the SBAS discovery papers, with a median difference of 2.5\%.    

	In order to try to quantify the uncertainty in our measurements, we measured the Einstein radii for all the objects again several months after the first measurement without referencing previous data. In all cases the differences between the original and new measurements were between $0.03\arcsec$ and $0.6\arcsec$. Since this represents up to $10\%$ of the value of $\theta_E$, we estimated the uncertainty in $\theta_E$ as $10\%$.  

	We note however that this method of estimating Einstein radius can lead to large systematic errors, so we also compared our values for Einstein radii to values from other groups.  \citet{West12} present strong lensing models for three of our systems and \citet{Oguri12} present models for four of our systems.  Both groups have measurements for SDSS J1343+4155, so we compared values for a total of six systems.  We provide measured Einstein radii from these papers in Table \ref{Einscompare}.  For SDSS J0900+2234 and SDSS J0901+1814, our estimates are almost exactly the same as the values in \citet{West12}.  However for the other four systems, the scatter (standard deviation) in values is larger, between $2.1\arcsec$ and $4.0\arcsec$.  We account for this error by calculating the fractional error in the values for $\theta_E$ and then finding the median value of the fractional errors for each of the six systems.  The median value of the fractional errors is $0.32$, or $32\%$, which we added in quadrature to the $10\%$ errors to find final error values.

	Solving Equation(\ref{eqn_eins}) for the mass, we obtain:
\begin{equation}
{M=\theta_E^2{\frac{c^2}{4G}\frac{D_{d}D_{s}}{D_{ds}}}}
\end{equation}	
Using the redshifts listed in Table \ref{redshifts} for the galaxy clusters and the source galaxies, we calculated the angular diameter distances.  We then used the Einstein radii we had measured to calculate the masses of the lenses. 

	Finally, we calculated the velocity dispersions of the regions of the clusters inside $\theta_E$ assuming the mass distribution was well fit by a singular isothermal sphere (SIS).  We used the following equation, from \citet{Narayan97}:
\begin{equation}
\sigma_v={\sqrt{\frac{\theta_E c^2 D_{s}}{4\pi D_{ds}}}}        \label{velodisp}          
\end{equation}	
All values measured for the strong lenses are presented in Table \ref{lenses}.
	
\begin{figure}
\begin{center}
\includegraphics[scale=0.8, angle=90]{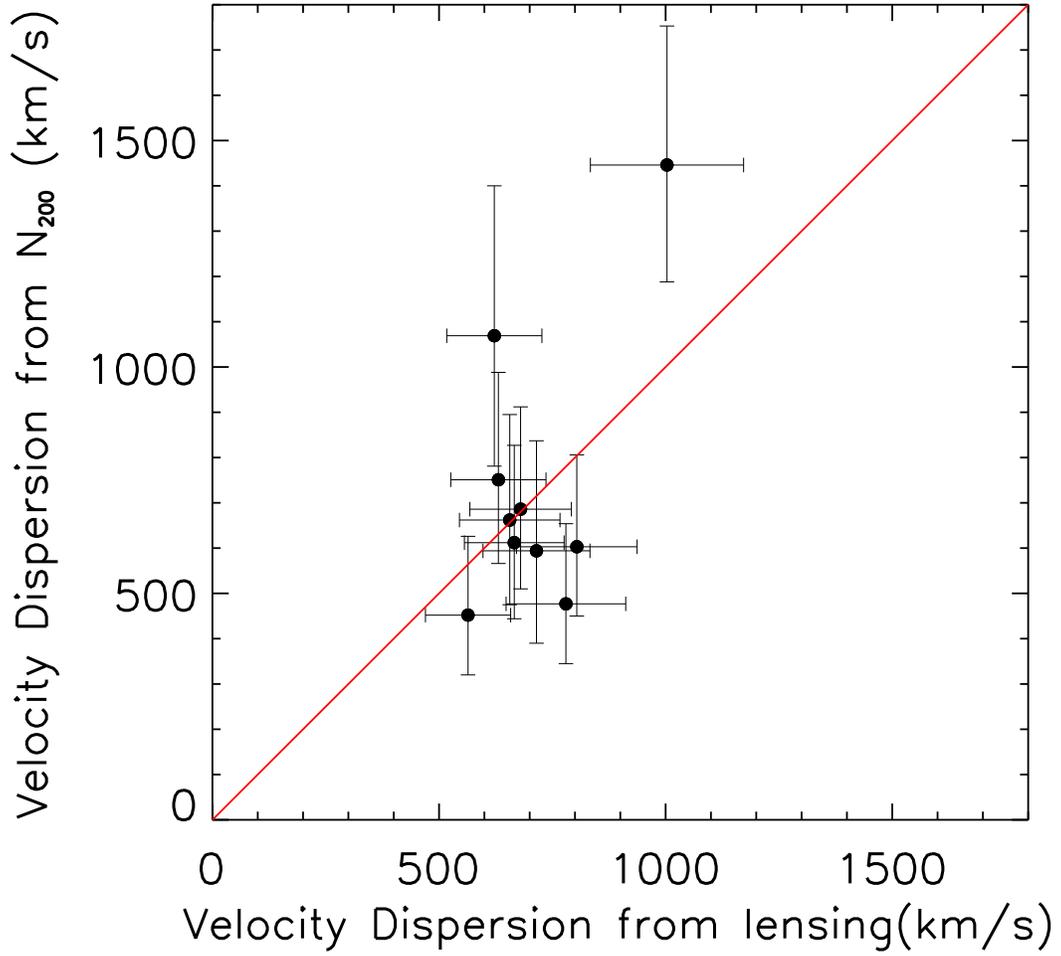}
\caption{A comparison of velocity dispersions found from $N_{200}$ and found from Einstein radii.  The line shown has the equation $y=x$.  The clusters on or above the $y=x$ line are all higher mass clusters.}\label{velocity}
\end{center}
\end{figure}

	In Figure \ref{velocity} we compare the velocity dispersions found from lensing to those found from richness measurements.  Note that these velocity dispersions measure different things:  the velocity dispersion from lensing describes the velocity dispersion inside $\theta_E$ and the velocity dispersion from $N_{200}$ describes the velocity dispersion within the much larger distance $r_{200}$.  We see in Figure \ref{velocity} that many of the clusters are found along the $y=x$ line, several are found above it and several are found below it.  For the clusters found along the $y=x$ line, we see that the velocity dispersions are similar within the two different radii, $\theta_E$ and $r_{200}$, which suggests that these systems are largely isothermal.   For the systems found above the $y=x$ line, the velocity dispersion at large radii is much larger than at small radii, indicating that much of the mass is found at larger distance from the BCG, suggesting a low value for $c_{200}$.  However for several of the clusters, the velocity dispersion within $\theta_E$ is larger than that found within $r_{200}$, indicating that for several clusters there is more mass within the smaller radius and suggesting that the concentration parameter is large.  Our highest mass clusters are found above the $y=x$ line (suggesting lower concentration parameter), while our lower mass clusters are found below the $y=x$ line (suggesting higher concentration parameter).  This would agree with what we discuss in the next section, that our highest mass clusters are not overconcentrated but our lowest mass clusters seem to be.
		
\section{Applications to Cosmology}
\subsection{An Overconcentration Problem?}   \label{overconc}
  Several recent papers \citep{OguriBland09,Gralla11, Fedeli11,Oguri12} have presented evidence that galaxy clusters that exhibit strong lensing have higher concentration parameters than $\Lambda$CDM would predict.  The most recent considerations \citep{Fedeli11,Oguri12} suggest that this overconcentration is most significant at cluster masses less than $10^{14}h^{-1}M_{\sun}$.  Overconcentration can be illustrated by comparing Einstein radii to $M_{200}$ \citep{Gralla11}.  Since Einstein radii are dependent on both cluster mass and cluster concentration parameter, such a comparison will yield larger Einstein radii than would be expected for particular $M_{200}$ values if the clusters are overconcentrated.

	Considering this, we have compared Einstein radius to $M_{200}$ for our ten systems. One complication in making this comparison is that Einstein radius is a function of redshift.  Since all of our systems have different redshifts for both lens and source, in order to compare them, we needed to scale them to a single, constant redshift for lens and source. We chose both the lens and source redshifts (we refer to them henceforth as fiducial redshifts) by taking the mean of the ten lens redshifts and the mean of the ten source redshifts. Our fiducial redshifts are $z_d=0.433$ for the lens and $z_s=1.65$ for the source.
	
	To scale Einstein radii to the fiducial redshifts, we needed to find a scale factor $k$ that would satisfy:	
\begin{equation}
\theta_{E\_scaled}(z_{d\_fiducial},z_{s\_fiducial})=k\times\theta_E\_{measured}(z_d,z_s)  \label{constant}
\end{equation} 	
We note that Equation \ref{velodisp} can be rearranged as
\begin{equation}
\theta_E=\frac{4\pi\sigma_{v}^2}{c^2}\frac{D_{ds}}{D_s}
\end{equation}	
Since $\sigma_v$ is proportional to the mass and does not depend on redshift, $\theta_E$ scales with redshift according to the ratio $D_{ds}/D_{s}$.  Thus solving Equation \ref{constant} for $k$ we obtain: 
\begin{equation}
k=\frac{\theta_E\_{scaled}}{\theta_E\_{measured}}=\frac{\frac{4\pi\sigma_{v}^2}{c^2}\frac{D_{ds\_fiducial}}{D_s\_fiducial}}{\frac{4\pi\sigma_{v}^2}{c^2}\frac{D_{ds}}{D_s}}   
\end{equation}	
and since $\sigma_v$ does not scale with redshift, it cancels.  Then
\begin{equation}
k=\frac{D_{ds\_fiducial}/{D_{s\_fiducial}}}{{D_{ds}}/{D_s}} \label{fiducial}
\end{equation}
We applied Equation \ref{fiducial} to find the scale factor $k$ for each cluster and then scaled each Einstein radius to the fiducial values.  

	In order to compare the relation between Einstein radius and $M_{200}$ for our data to the relation that $\Lambda$CDM would predict, we refer to the models presented in \citet{Oguri09} and \citet{Oguri12} which predict concentration as a function of cluster mass.  Concentration parameter, $c_{\Delta}$, is defined as
\begin{equation}
c_{\Delta}=\frac{r_{\Delta}}{r_s}
\end{equation}
The $r_s$ term is the scale radius, a term in the Navarro-Frenk-White (NFW) model of dark matter halo density \citep{NFW97}, described below.  The quantity $\Delta$ is the virial overdensity.  In this paper we use $\Delta=200$, but \citet{Oguri09} use $\Delta=vir$, where the virial overdensity is the local overdensity that would cause halo collapse; it is a function of redshift.  \citet{Oguri09} suggest that lensing-selected clusters (those discovered based on lensing, like those in this paper) will have a value for the concentration that is $50\%$ higher than for general clusters.  
	
	\citet{Oguri09} present a relation for $c_{vir}$ in general clusters, citing results obtained from N-body simulations conducted using WMAP5 cosmology \citep{Duffy08}: 
\begin{equation}
\bar{c}_{vir}(sim)=\frac{7.85}{(1+z)^{0.71}}\left(\frac{M_{vir}}{2.78\times10^{12}M_\sun }\right)^{-0.081} \label{Oguri1}
\end{equation}
We consider this relation at $z=0.45$, for consistency with the lensing-selected relation below.		
\citet{Oguri12} present a relation for $c_{vir}$ in lensing-selected clusters, using ray tracing to estimate the effect of lensing bias:
\begin{equation}
\bar{c}_{vir}(z=0.45)\approx6.3\left(\frac{M_{vir}}{5\times10^{14}h^{-1}M_{\sun}}\right)^{-0.2} \label{Oguri2}
\end{equation}

	In order to compare our data to these predictions, we chose a range of values of $M_{vir}$ and used Equations \ref{Oguri1} and \ref{Oguri2} to find the corresponding values for $c_{vir}$.  We then used the relations in \citet{Johnston07} and \citet{HuKravtsov03} to convert from $c_{vir}$ and $M_{vir}$ to $c_{200}$ and $M_{200}$.  Finally we used the range of values for $M_{200}$ and the predicted values for $c_{200}$ to find predicted values for Einstein radius ($\theta_E)$ by using the NFW profile (see Equation \ref{Navarro} below).  We plotted the relations between $M_{200}$ and $\theta_E$ as the general and lensing-selected predictions in Figures \ref{dispersions_old} and \ref{dispersions}. 

	To find a predicted Einstein radius we used the NFW density profile, expressed as
\begin{equation}
\rho(r)=\frac{\rho_s}{\left(r/r_s\right)\left(1+r/r_s\right)^2}  \label{Navarro}
\end{equation}	
where $r$ is the distance from the center of the cluster, $\rho_s$ is a characteristic density, and $r_s$ is the scale radius, given by $r_s = r_{200}/ c_{200}$.  We implemented Equation 13 in \citet{Wright00}, an equation that describes surface mass density $\overline{\Sigma}_{NFW}$ in the NFW model.  The Einstein radius $\theta_E$ is given implicitly by the solution of \citep{Narayan97}:
\begin{equation}
\overline{\Sigma}_{NFW}\left(\frac{\theta_{E}}{r_s}\right)=\Sigma_{crit} \label{crit}
\end{equation}
where the critical surface mass density $\Sigma_{crit}$ is 
\begin{equation}
\Sigma_{crit}=\frac{c^2}{4\pi G}\frac{D_s}{D_{d}D_{ds}}
\end{equation}
Thus we found Einstein radius by solving for $\overline{\Sigma}_{NFW}$ and using that to find $\theta_{E}$. 

\subsection{Consideration of the Overconcentration Problem} \label{final}
\begin{figure}
\includegraphics[scale=0.8, angle=90]{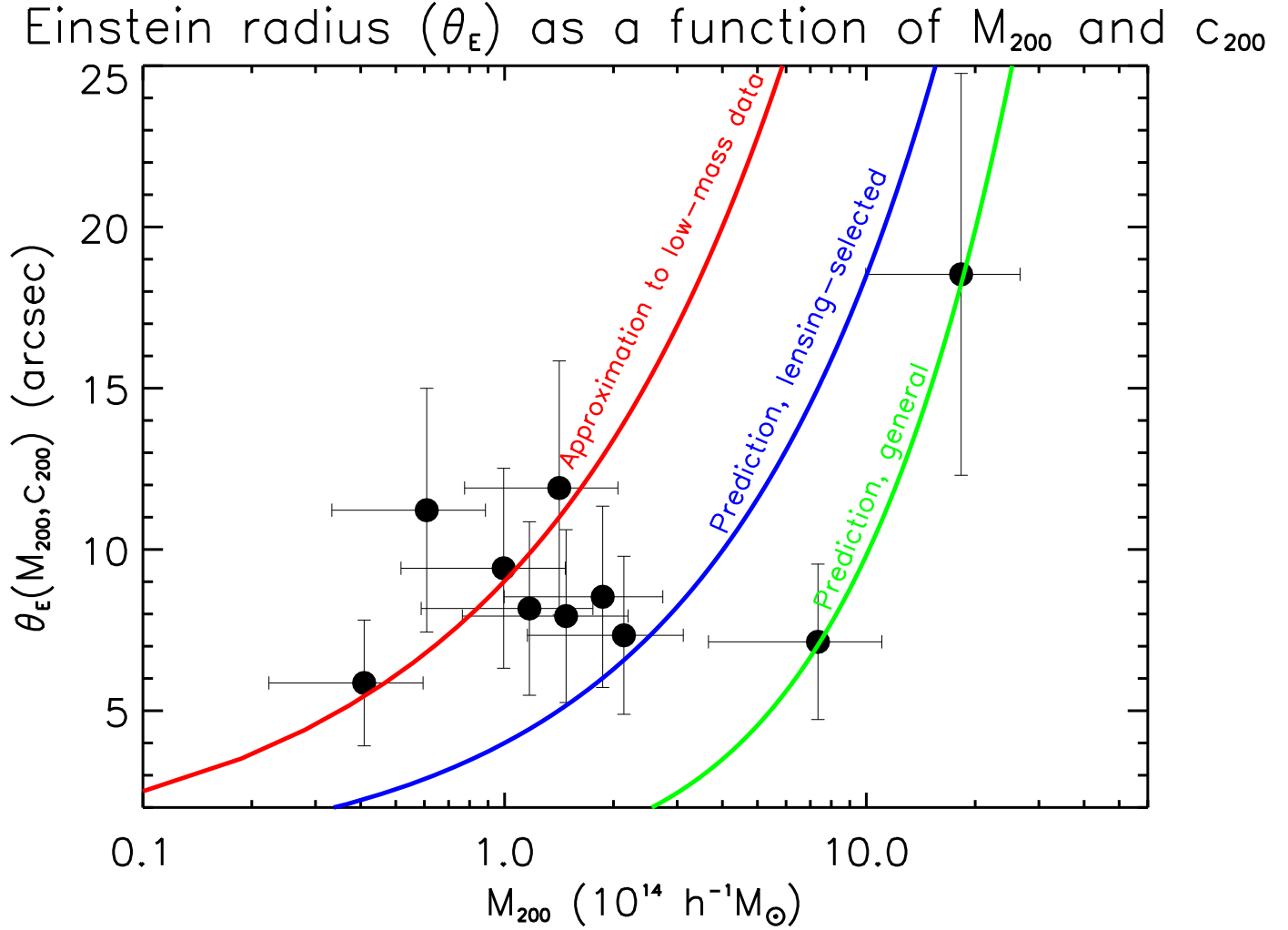}
\caption{A plot of Einstein radius versus $M_{200}$ for unscaled $M_{200}$ values, with Einstein radii scaled to fiducial redshifts.  The theoretical lines come from a prediction of Einstein radii for given $M_{200}$ and $c_{200}$ values found by using an NFW \citep{NFW97} fit to the mass and concentration.  The general clusters line was found using predicted $c_{200}$ values found from Equation \ref{Oguri1} and the lensing-selected clusters line was found using predicted $c_{200}$ values found from Equation \ref{Oguri2}.  The average $c_{200}$ for general clusters is $3.6$ and for lensing-selected clusters it is $6.4$.  Both Equations took $z=0.45$.  The approximate fit line was found by multiplying the values of $c_{vir}$ resulting from Eq. \ref{Oguri2} by $1.9$.  We tried different factors to multiply $c_{vir}$ until the resultant line went approximately through the low mass data points.}\label{dispersions_old}
\end{figure}

\begin{figure}
\includegraphics[scale=0.8, angle=90]{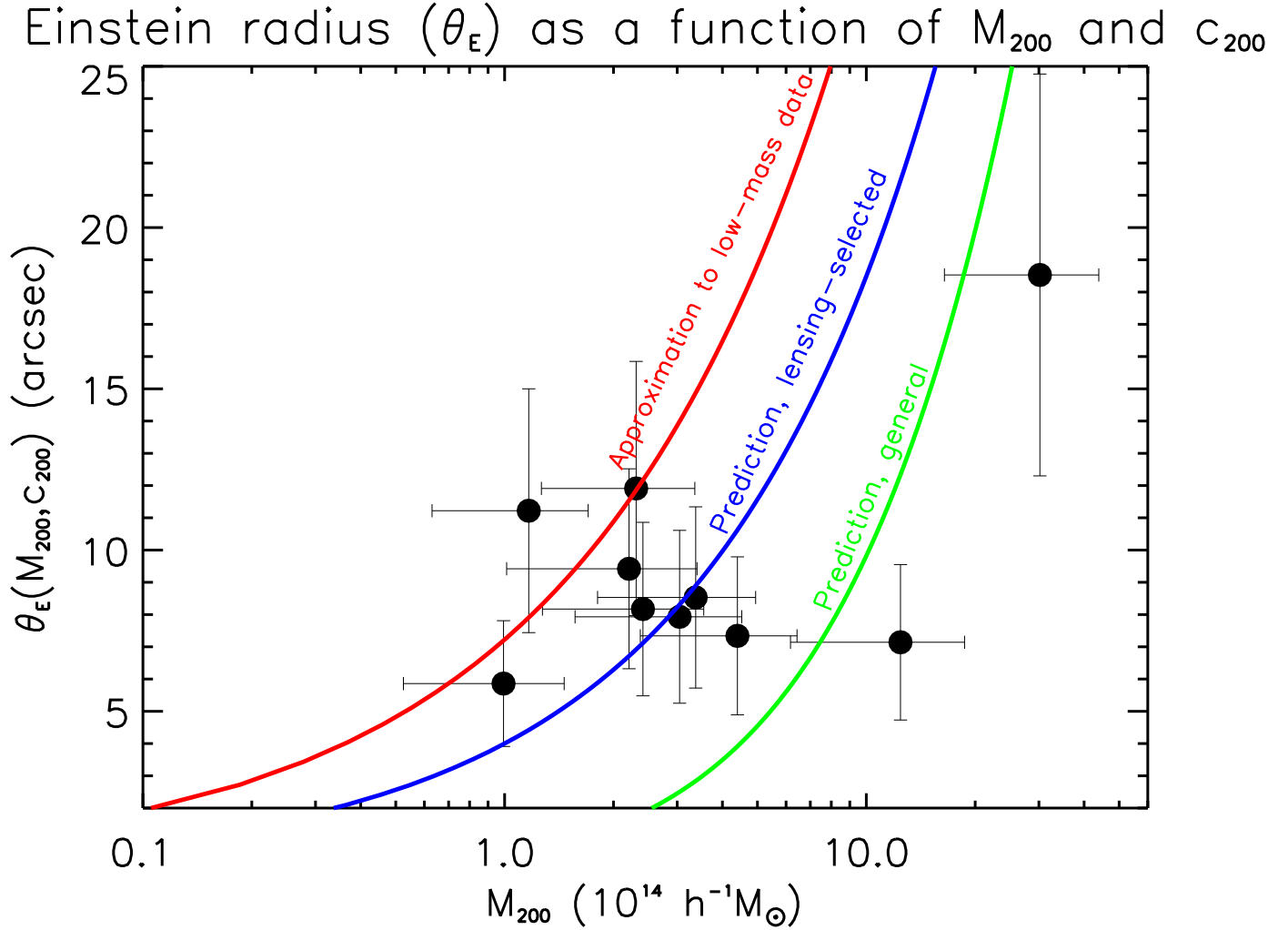}
\caption{The same plot as Figure \ref{dispersions_old} but using $M_{200}$ values found by scaling richness values up using Eq. \ref{scaler}.  The average $c_{200}$ for general clusters is $3.4$ and for lensing-selected clusters it is $5.7$.  To find the approximate fit to the low mass data, we multiplied all $c_{vir}$ values from Eq. \ref{Oguri2} by $1.5$.  Note that with scaled richness values, the points all move closer to the predicted values.}\label{dispersions}
\end{figure}

	The final result of our analysis is shown in Figures \ref{dispersions_old} and \ref{dispersions}.  Figure \ref{dispersions_old} shows the relation between $M_{200}$ and $\theta_E$ for our measured values of $M_{200}$ while Figure \ref{dispersions} shows the relation for the new $M_{200}$ values that come from the scaled-up richness values.  We consider Figure \ref{dispersions} to be more reliable as it uses richness values scaled to correspond with values from SDSS data, which was used to calibrate the mass-richness relation.  In Figure \ref{dispersions_old} there is a noticeable disagreement between our data and the predicted relations.  It can also be seen that the lower-mass clusters disagree more while the higher-mass clusters fit the predictions better, as found by other authors.  However in Figure \ref{dispersions}, we see that all clusters are shifted to higher masses by an average factor of $2.0$.  In the plot of the scaled values, we see that many of the clusters now closely follow the lensing-selected prediction.  There are still four clusters that do not fit the predicted relations.  These clusters are SDSS J0901+1814, SDSS J1038+4849, SDSS J1343+4155 and SDSS J1537+6556, which are the lowest mass clusters in our sample.  SDSS J1318+3942, which is also among the lowest mass clusters, is found close to the predicted line, but still slightly above it.   

	We determined values for $c_{200}$ for our clusters by using our measured values for $M_{200}$ and $\theta_E$ in Equations \ref{Navarro} and \ref{crit}; values are listed in Table \ref{clusters}.  We estimated errors on $c_{200}$ by varying $M_{200}$ and $\theta_E$ to the maximum and minimum values allowed by their respective error bars.  Maximum values for $c_{200}$ were found with minimum $M_{200}$ and maximum $\theta_E$ while minimum values for $c_{200}$ were found with the opposite.  For smaller values of $M_{200}$, this led to very large upper error bars on $c_{200}$ as a very high concentration parameter would then be required to achieve the large Einstein radius.  
	
	Our measurements of $c_{200}$ follow the trends noted earlier:  for many of the clusters, our measured values of $c_{200}$ are within the range of predictions, but for the lowest mass clusters measured values of $c_{200}$ are higher than predictions.  The average value for $c_{200}$ predicted for our scaled values of $M_{200}$ by Equation \ref{Oguri1} (for general clusters) is $3.4$ while the average value predicted by Equation \ref{Oguri2} (for lensing-selected clusters) is $5.7$.  The average of our ten measured values of $c_{200}$ is $7.3$, which is slightly larger than the lensing-selected prediction.  However for our four lowest mass clusters the average $c_{200}$ value is $11.6$, much larger than the lensing-selected prediction.  The four clusters we identify as overconcentrated above have the following values for $c_{200}$:  for SDSS J0901+1814 $c_{200}=9.6_{-3.5}^{+13}$, for SDSS J1038+4840 $c_{200}=17_{-7.8}^{+73},$ for SDSS J1343+4155 $c_{200}=9.1_{-3.9}^{+22}$ and for SDSS J1537+6556, $c_{200}=11_{-3.9}^{+15}$.  These clusters have respectively $M_{200}$ of $0.99$, $1.2$, $2.3$ and $2.2\times10^{14} h^{-1}M_{\sun}$, which are the lowest masses in our sample.
	
	Concentration parameters ($c_{vir}$) based on strong and weak lensing measurements are provided in \citet{Oguri12} for two of these four clusters.  We convert these to $c_{200}$ using the method discussed in \S \ref{overconc}.  For SDSS J1038+4840, $c_{200}=33.8_{-18.3}^{+0.00}$ and for SDSS J1343+4155, $c_{200}=4.25_{-0.790}^{+1.38}$.  Thus for SDSS J1038+4840, the second lowest mass cluster in our sample, both sets of measurements find this cluster to be significantly overconcentrated.  For SDSS J1343+4155 the evidence for overconcentration is not as strong.
	
\begin{figure}
\includegraphics[scale=0.8, angle=90]{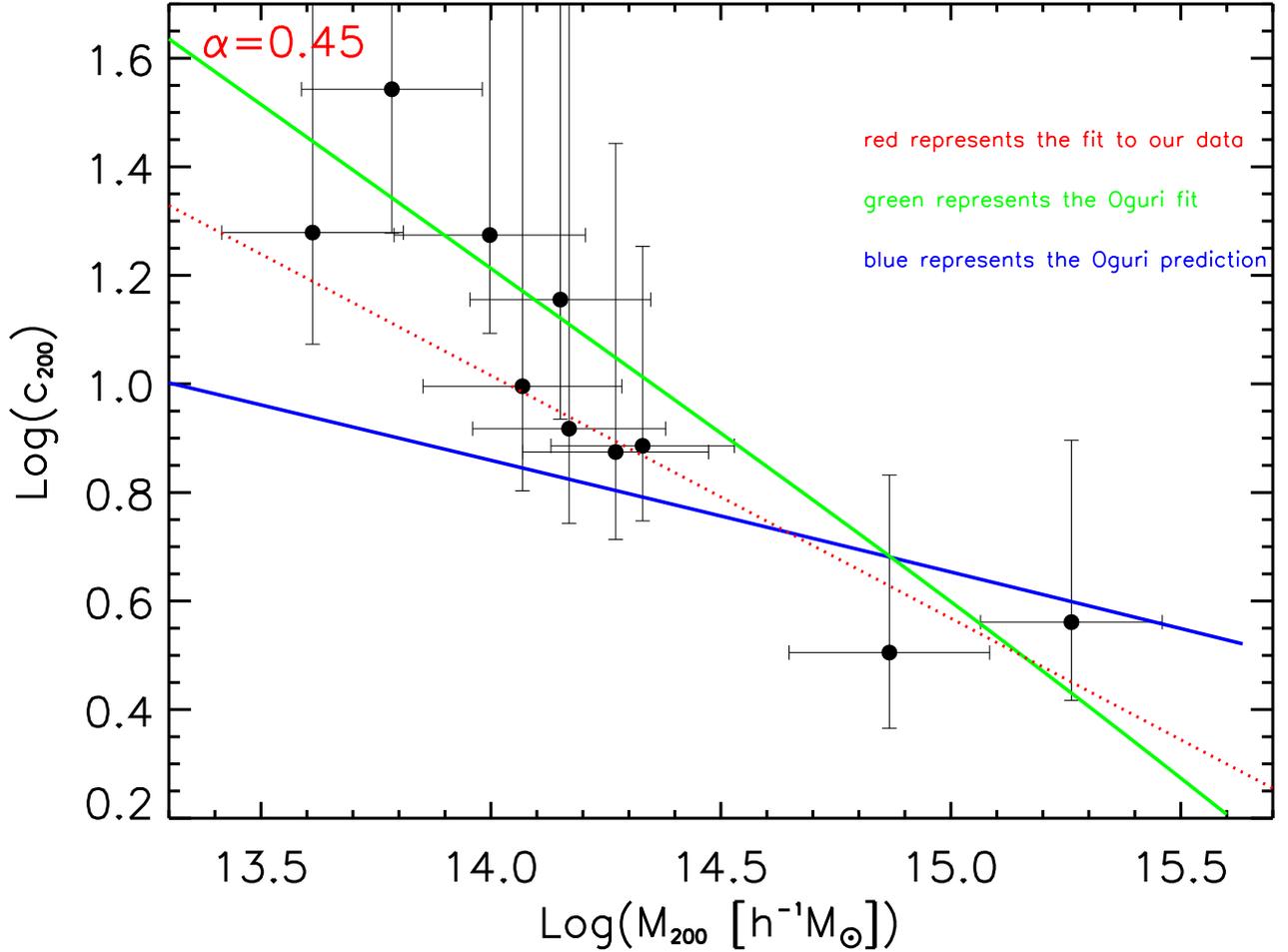}
\caption{A plot of the logarithm of the concentration parameter $c_{200}$ versus logarithm of $M_{200}$.  This is modeled on Figure 1 in \citet{Fedeli11}.  The dotted red line is the best fit to the data and has a slope $\alpha = 0.45 \pm 0.30$.  The solid green line (with the higher slope) is the fit to the data in \citet{Oguri12} (their Equation 26).  The solid blue line (with the lower slope) is the Oguri prediction for lensing-selected clusters (Equation \ref{Oguri2}).  The large vertical error bars arise on the low-mass clusters due to how $c_{200}$ changes as a function of $M_{200}$ and $\theta_E$.   We found the upper vertical error bars on $c_{200}$ by setting $M_{200}$ and $\theta_{E}$ to their minimum and maximum values, respectively.  When $M_{200}$ is very small, a very large value for $c_{200}$ is required to achieve the large value for Einstein radius.} \label{Fedeli_old}       
\end{figure}

\begin{figure}
\includegraphics[scale=0.8, angle=90]{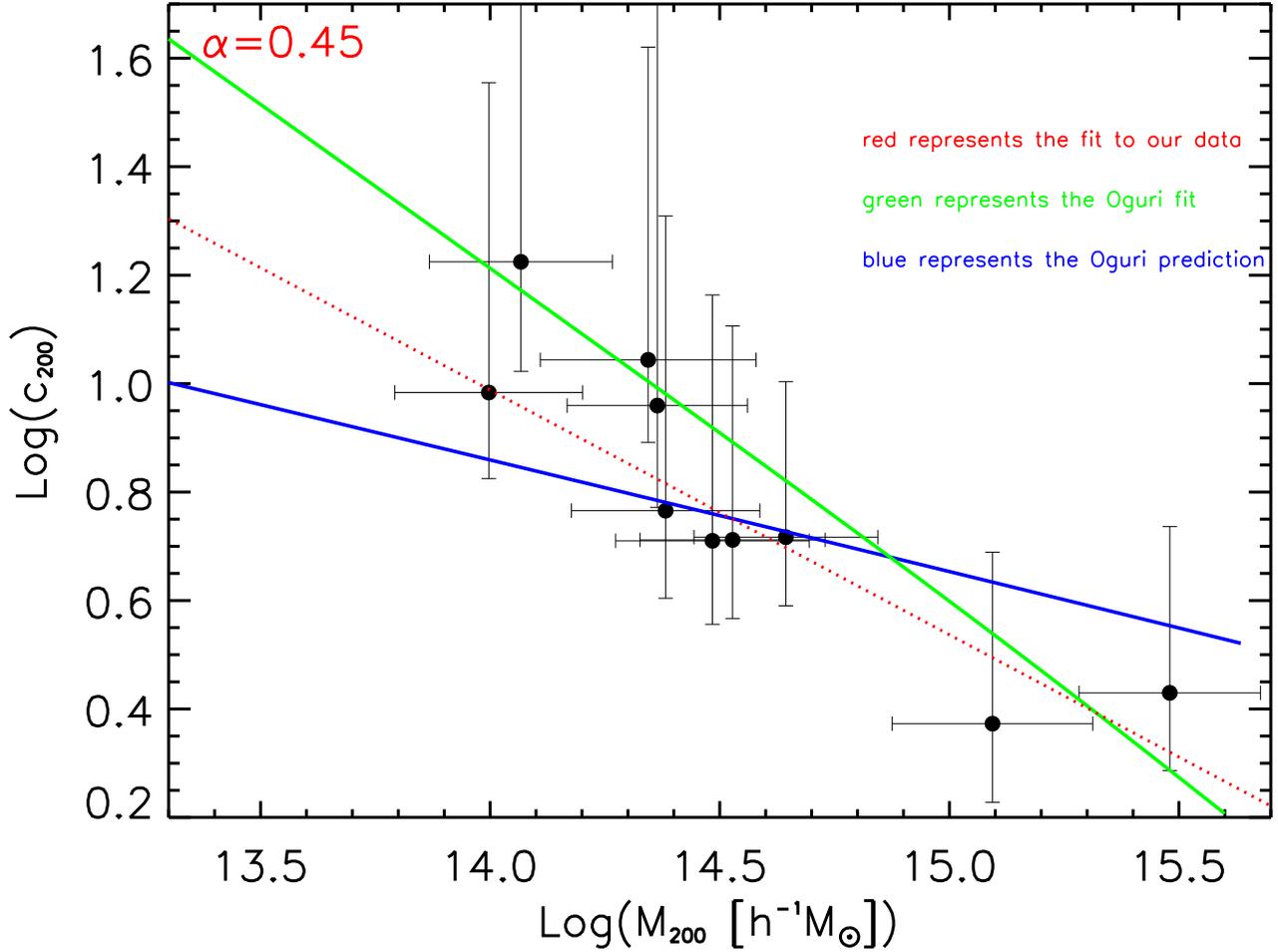}
\caption{The same plot as Figure \ref{Fedeli_old} but made with $M_{200}$ and $c_{200}$ values corresponding to richness values scaled up using Eq. \ref{scaler}.  The slope of the best fit line is $\alpha=0.45 \pm 0.23$.  Note that the values for $M_{200}$ have been shifted to the right and thus many of the points fit the predicted relations now.  However the lowest mass points still do not match the predictions.} \label{Fedeli}       
\end{figure}

	In Figures \ref{Fedeli_old} and \ref{Fedeli} we consider the mass-concentration relation, comparing log($c_{200}$) to log($M_{200}$).  Figure \ref{Fedeli_old} is the mass-concentration relation for our measured values of $c_{200}$ and $M_{200}$ found without scaling and Figure \ref{Fedeli} is this relation for values found with scaling.  We also include three lines in Figures \ref{Fedeli_old} and \ref{Fedeli}:  the blue solid line is the prediction from \citet{Oguri12} for lensing-selected clusters, the green solid line is the best-fit to the data in the Oguri paper (Equation 26 in \citet{Oguri12}) and the red dotted line is the best fit to our data.  Equation 26 in \citet{Oguri12} is:
\begin{equation}
\overline{c_{vir}}=(7.7 \pm 0.6)\left(\frac{M_{vir}}{5\times10^{14}h^{-1}M_{\sun}}\right)^{-0.59 \pm 0.12} \label{26}
\end{equation}
We used the same method as discussed in \S \ref{overconc} to add the prediction and best fit from \citet{Oguri12} to Figures \ref{Fedeli_old} and \ref{Fedeli}.  For the predicted line, we applied Equation \ref{Oguri2} and for the best fit from \citet{Oguri12} we applied Equation \ref{26}.  In Figure \ref{Fedeli_old} the slope is $\alpha=0.45 \pm 0.30$ while in Figure \ref{Fedeli} $\alpha=0.45 \pm 0.23$.  Note that the error bars are larger on $c_{200}$ in Figure \ref{Fedeli_old}; this is because when calculating error bars, the minimum $M_{200}$ was small and maximum $\theta_E$ was large, leading to very large values for $c_{200}$.  \citet{Fedeli11} suggests that for clusters that are not overconcentrated, $\alpha$ should be no larger than $0.2$.  At $1\sigma$, our lowest value of $\alpha$ is $0.15$ for unscaled values and $0.22$ for scaled values.  Both of these values are consistent with clusters that are not overconcentrated, again suggesting that most of our clusters are not overconcentrated.  \citet{Prada11} suggest in their Figure 12 that log($c_{200}$) should be less than about $0.8$ at $z=0.5$.  This is again consistent with most of our clusters, although not for the lowest mass clusters.  Note in Figure \ref{Fedeli} that the four lowest mass clusters have values of log($c_{200})$ above $1.0$ which suggest that these clusters are overconcentrated.

	We find in Figure \ref{Fedeli_old} that our data points are mostly above the predicted line, suggesting many of our clusters are overconcentrated.  However when we use the more reliable scaled values in Figure \ref{Fedeli} we find that most of the clusters are found near the predicted line, but the lowest mass clusters (the four identified above) remain above the prediction.  This again confirms our previous statement that most of our clusters do not appear to be overconcentrated, but there is evidence for overconcentration at lower cluster masses.  	
	
	Thus for most of our clusters, $\Lambda$CDM seems to match their observed properties.  But for our several clusters showing evidence of overconcentration, what does the overconcentration problem suggest is happening in galaxy clusters?  It seems to suggest that clusters are collapsing more than $\Lambda$CDM would predict \citep{Broad08,Fedeli11,Oguri12}.  The dark matter halo associated with a galaxy cluster is expected to have undergone an adiabatic collapse during the formation of the cluster.  The baryonic matter in the cluster (concentrated in the BCG) would also have collapsed.  The baryonic matter would likely have dragged the dark matter along with it, augmenting the collapse of the halo.  Since we find some clusters to be more concentrated than expected, it may be that the halo collapsed more than expected due to the contribution of the baryons.  It is suggested \citep{Fedeli11,Oguri12} that the overconcentration is most significant in lower-mass clusters because in these clusters the BCG makes up a larger percentage of the overall cluster mass.  Thus the baryons would contribute to the halo collapse more in a lower-mass cluster than in a higher-mass cluster.
	
\section{Conclusion}
	We have reported on the properties of ten galaxy clusters exhibiting strong gravitational lensing arcs which were discovered in the Sloan Digital Sky Survey.  These are a subset of the $19$ systems discovered thus far by the Sloan Bright Arcs Survey.  
	
	We measured $N_{200}$, $M_{200}$, $\sigma_{v}$ and $c_{200}$ using the postulates of the maxBCG method to identify cluster galaxies.  We found that the values of $N_{200}$ measured here do not agree with values found in the SDSS.  This is because magnitude errors are larger in SDSS data so some non-cluster galaxies are scattered into the sample, overestimating cluster richnesses.  Thus we scaled our $N_{200}$ values up to match the SDSS values in order that we might use the mass-richness relation calibrated from the SDSS.  The scaled richness values for the clusters range from $N_{200}=22$ to $N_{200}=317$.  The cluster masses range from M$_{200}=0.993\times10^{14}h^{-1}M_{\sun}$ to M$_{200}=30.2\times10^{14}h^{-1}M_{\sun}$ and the velocity dispersions for the clusters range from $\sigma_v=452$ km/s to $\sigma_v=1446$ km/s.  Finally the concentration parameters for the clusters range from $2.4$ to $17$.  
	
	We applied a simple SIS model to infer the lens masses and lens velocity dispersions from the measured Einstein radii. The smallest Einstein radius was $\theta_E=5.4\arcsec$ and the largest was $\theta_E=13\arcsec$.  The lens mass within the Einstein radius ranged from $M=5.5\times10^{12}h^{-1}M_{\sun}$ to $M=36\times10^{12}h^{-1}M_{\sun}$ and the lens velocity dispersion ranged from $\sigma_v=336$ km/s to $\sigma_v=804$ km/s. 
	
	Finally we considered the relation between $\theta_E$ and $M_{200}$ and compared this relation to the predictions of $\Lambda$CDM, both for lensing-selected and for general clusters.  We also found the mass-concentration relation for our data.  We found that most of our clusters are not overconcentrated, but our four lowest mass clusters show evidence of overconcentration, with values for $c_{200}$ between $9.6$ and $17$.  This may suggest that the lowest mass clusters are collapsing more than $\Lambda$CDM would predict.
	
\acknowledgments
	Funding for the SDSS and SDSS-II has been provided by the Alfred P. Sloan Foundation, the Participating Institutions, the National Science Foundation, the U.S. Department of Energy, the National Aeronautics and Space Administration, the Japanese Monbukagakusho, the Max Planck Society, and the Higher Education Funding Council for England. The SDSS Web Site is http://www.sdss.org/.

	The SDSS is managed by the Astrophysical Research Consortium for the Participating Institutions. The Participating Institutions are the American Museum of Natural History, Astrophysical Institute Potsdam, University of Basel, University of Cambridge, Case Western Reserve University, University of Chicago, Drexel University, Fermilab, the Institute for Advanced Study, the Japan Participation Group, Johns Hopkins University, the Joint Institute for Nuclear Astrophysics, the Kavli Institute for Particle Astrophysics and Cosmology, the Korean Scientist Group, the Chinese Academy of Sciences (LAMOST), Los Alamos National Laboratory, the Max-Planck-Institute for Astronomy (MPIA), the Max-Planck-Institute for Astrophysics (MPA), New Mexico State University, Ohio State University, University of Pittsburgh, University of Portsmouth, Princeton University, the United States Naval Observatory, and the University of Washington.
	
	This paper is based on observations obtained at Kitt Peak National Observatory, which is operated by the Association of Universities for Research in Astronomy, Inc., under a cooperative agreement with the NSF.
	
	S. S. Allam acknowledges support from an HST Grant. Support of program no. 11167
was provided by NASA through a grant from the Space Telescope Science Institute, which
is operated by the Association of Universities for Research in Astronomy, Inc., under NASA
contract NAS5-26555.

Fermilab is operated by Fermi Research Alliance, LLC under Contract No. DE-AC02-
07CH11359 with the United States Department of Energy.

M. P. Wiesner would like to acknowledge guidance and support from M. Fortner, L. Lurio and everyone in the Department of Physics at NIU.

\begin{center}
\begin{table}
 \small
    \begin{tabular}{c  c  c  c  c}\tableline\tableline
 \bf{System} & \bf{R.A. (deg)}& \bf{Decl. (deg)} & \bf{Lens z} & \bf{Source z}\\ \tableline
     SDSS J0900+2234&135.01128&22.567767&0.4890&2.0325\\ 
     SDSS J0901+1814&135.34312&18.242326&0.3459&2.2558\\ 
     SDSS J0957+0509&149.41318&5.1589174&0.4469&1.8230\\ 
     SDSS J1038+4849&159.67974&48.821613&0.4256&0.966\\ 
     SDSS J1209+2640&182.34866&26.679633&0.5580&1.018\\ 
     SDSS J1318+3942&199.54798&39.707469&0.4751&2.9437\\ 
     SDSS J1343+4155&205.88702&41.917659&0.4135&2.0927\\ 
     SDSS J1439+3250&219.98542&32.840162&0.4176&1.0-2.5\tablenotemark{a}\\ 
     SDSS J1511+4713&227.82802&47.227949&0.4517&0.985\\ 
     SDSS J1537+6556&234.30478&65.939313&0.2595&0.6596\\ \tableline
     \end{tabular}
     \tablenotetext{a}{Source redshift has not yet been determined for this system, thus we present a range of possible values.}
\caption{The coordinates and redshifts of the ten systems in this paper.}\label{redshifts}
\end{table}
\end{center}

\begin{center}
\begin{table}
\tablewidth{0pt}
\small
\begin{tabular}{c c c c c}\tableline \tableline
\bf{System} & \bf{$\Delta$m} & \bf{g-r color} & \bf{r-i color} & \bf{0.4L* Magnitude} \\ \tableline
SDSS J0900+2234&0.56&1.83&0.73&21.20\\
SDSS J0901+1814&0.22&1.72&0.52&20.26 \\
SDSS J0957+0509&0.15&1.78&0.71&21.26 \\
SDSS J1038+4849&0.07&1.72&0.62&20.84\\
SDSS J1209+2640&0.34&1.79&0.93&21.59\\
SDSS J1318+3942&0.06&1.73&0.73&21.15 \\
SDSS J1343+4155&0.16&1.75&0.54&20.71\\
SDSS J1439+3250&0.11&1.74&0.67&20.78\\
SDSS J1511+4713&0.17&1.78&0.75&20.97 \\
SDSS J1537+6556&0.14&1.50&0.52&19.38 \\ \tableline
     \end{tabular}
   \caption{A summary of the values of limits used for richness measurements.  $\Delta$m is the magnitude measured in i-band in 2$\arcsec$ $\tt{MAG\_APER}$ minus the magnitude in the same band measured in $\tt{MAG\_AUTO}$.  $\Delta$m was used for star-galaxy separation.  The $g-r$ and $r-i$ colors are based on measurements in the 2$\arcsec$ aperture.  Finally, the magnitude at $0.4L^*$ was found in the i-band.}\label{colors}
\end{table}
\end{center} 

\begin{center}
\begin{table}
\tablewidth{0pt}
\small
\begin{tabular}{c c c c c}\tableline \tableline
\bf{System} &${N_{gals}}$ ($\tt{MAG\_AUTO}$)& ${N_{gals}}$ $(2\arcsec \ \tt{MAG\_APER})$&${N_{gals}}$ $(3\arcsec \ \tt{MAG\_APER})$&
${N_{gals} (Sloan)}$\\ \tableline
SDSS J0900+2234&23&28&29&56\\ 
SDSS J0901+1814&8&14&8&11\\
SDSS J0957+0509&15&28&26&63\\ 
SDSS J1038+4849&16&15&17&32\\ 
SDSS J1209+2640&85&101&98&190\\ 
SDSS J1318+3942&21&23&23&39\\ 
SDSS J1343+4155&26&25&32&46\\
SDSS J1439+3250&48&55&51&82\\ 
SDSS J1511+4713&22&29&29&54\\
SDSS J1537+6556&14&20&18&8\\  \tableline
     \end{tabular}
   \caption{A comparison of $N_{gals}$ values measured in different SExtractor apertures and in SDSS data.  These values for $N_{gals}$ have not been area corrected with Eq. \ref{extrapolate}.  For the SDSS values, any area which is not on the CCD in the WIYN data is excluded from consideration.  Note that for SDSS J1537+6556 much of the WIYN area is outside the SDSS footprint, so the SDSS value is biased low.  \label{richvals}}
\end{table}
\end{center} 

\begin{center}
\begin{table}
\tablewidth{0pt}
\small
\begin{tabular}{c c c c c c c c c}\tableline \tableline
{System} & ${N_{gals}}$ & ${r_{200}(h^{-1}Mpc)}$& ${N_{200}}$& ${M_{200}(10^{14}M_\sun)}$& ${\sigma_v(km/s)}$ & ${c_{200}}$ \\ \tableline
SDSS J0900+2234&28&1.15&30 $\pm$ 4.1&1.48 $\pm$ 0.715&518$_{-153}^{+196}$&8.27$_{-3.32}^{+14.9}$	\\
SDSS J0901+1814&15&0.792&11 $\pm$ 0.58&0.409 $\pm$ 0.186&334$_{-98}^{+137}$&19.0$_{-9.00}^{+91.0}$ \\
SDSS J0957+0509&29&1.18&36 $\pm$ 3.4&1.87 $\pm$ 0.8708&561$_{-153}^{+200}$&7.49$_{-2.78}^{+9.81}$ \\
SDSS J1038+4849&16&0.823&15 $\pm$ 0.62&0.609 $\pm$ 0.276&383$_{-108}^{+150}$&34.9$_{-21.3}^{+18800}$ \\
SDSS J1209+2640&101&2.49&214 $\pm$ 11.5&18.3 $\pm$ 8.32&1219$_{-240}^{+293}$&3.64$_{-1.21}^{+2.81}$\\
SDSS J1318+3942&24&1.050&25 $\pm$ 4.2&1.17 $\pm$ 0.583&478$_{-150}^{+191}$&9.9$_{-4.39}^{+31.8}$ \\
SDSS J1343+4155&28&1.15&29 $\pm$ 1.1&1.42 $\pm$ 0.641&510$_{-135}^{+182}$&14.3$_{-7.26}^{+118}$ \\
SDSS J1439+3250&59&1.80&105 $\pm$ 18&7.35 $\pm$ 3.69&894$_{-250}^{+296}$&3.20$_{-1.03}^{+2.41}$ \\
SDSS J1511+4713&31&1.22&40 $\pm$ 2.9&2.14 $\pm$ 0.981&587$_{-154}^{+203}$&7.69$_{-2.45}^{+6.51}$ \\
SDSS J1537+6556&22&0.997&22 $\pm$ 2.9&0.994 $\pm$ 0.477&452$_{-136}^{+177}$&18.8$_{-7.84}^{+49.2}$ \\ \tableline
     \end{tabular}
   \caption{A summary of the quantities measured for the ten galaxy clusters without scaling.  These are all based on colors measured in the 2$\arcsec$  aperture.  The $N_{gals}$ and $N_{200}$ values are area-corrected using Eq. \ref{extrapolate} but are not scaled up.  All the other values are based on these area-corrected but not scaled richness values.}  \label{clusters}
\end{table}
\end{center} 

\begin{center}
\begin{table}
\tablewidth{0pt}
\small
\begin{tabular}{c c c c c c c c c}\tableline \tableline
{System} & ${N_{gals}}$ & ${r_{200}(h^{-1}Mpc)}$& ${N_{200}}$& ${M_{200}(10^{14}M_\sun)}$& ${\sigma_v(km/s)}$ & ${c_{200}}$ \\ \tableline
SDSS J0900+2234&28&1.45&53 $\pm$7.6&3.046 $\pm$ 1.48&662$_{-187}^{+233}$&5.13$_{-1.82}^{+5.36}$ \\
SDSS J0901+1814&15&0.996&22 $\pm$ 2.4&0.993 $\pm$ 0.468&452$_{-132}^{+174}$&9.63$_{-3.52}^{+12.7}$ \\
SDSS J0957+0509&29&1.48&57 $\pm$ 5.1&3.38 $\pm$ 1.57&686$_{-176}^{+226}$&5.15$_{-1.72}^{+4.68}$ \\
SDSS J1038+4849&16&1.036&25 $\pm$ 1.8&1.17 $\pm$ 0.536&477$_{-132}^{+177}$&16.8$_{-7.80}^{+73.3}$ \\
SDSS J1209+2640&101&3.13&317 $\pm$ 17&30.2 $\pm$ 13.7&1446$_{-258}^{+307}$&2.69$_{-0.890}^{+1.90}$ \\
SDSS J1318+3942&24&1.32&44 $\pm$ 5.0&2.41 $\pm$ 1.14&612$_{-168}^{+215}$&5.83$_{-2.17}^{+7.30}$ \\
SDSS J1343+4155&28&1.45&43 $\pm$ 1.6&2.31 $\pm$ 1.046&603$_{-153}^{+203}$&9.11$_{-3.94}^{+22.0}$ \\
SDSS J1439+3250&59&2.27&158 $\pm$ 28&12.4 $\pm$ 6.26&1069$_{-288}^{+331}$&2.36$_{-0.790}^{+1.72}$ \\
SDSS J1511+4713&31&1.54&70 $\pm$ 5.6&4.40 $\pm$ 2.030&751$_{-185}^{+237}$&5.21$_{-1.52}^{+3.44}$ \\
SDSS J1537+6556&22&1.25&41 $\pm$ 9.6&2.21 $\pm$ 1.20&594$_{-204}^{+243}$&10.0$_{-3.88}^{+14.7}$ \\ \tableline
     \end{tabular}
   \caption{A summary of the quantities measured for the ten galaxy clusters using the scaling described by Eq. \ref{scaler}.  These are all based on colors measured in the 2$\arcsec$  aperture.  The $N_{gals}$ and $N_{200}$ values are area-corrected using Eq. \ref{extrapolate}.}\label{scaled}
\end{table}
\end{center}

\begin{center}
\begin{table}
 \small
    \begin{tabular}{c c c c c}\tableline\tableline
 $\bf{System}$ & $\bf{M_{200}(this \ paper)}$& $\bf{M_{200}(Oguri)}$& $\bf{M_{200}(Bayliss)}$&$\bf{M_{200}(Drabek)}$ \\ \tableline
     SDSS J0957+0509&3.38 $\pm$ 1.57&1.17$_{-0.55}^{+0.77}$&8.01$_{-6.40}^{+4.98}$&-\\ 
     SDSS J1038+4849&1.17 $\pm$ 0.536&0.681$_{-0.11}^{+0.48}$&2.06$_{-0.36}^{+1.18}$&-\\ 
     SDSS J1209+2640&30.2 $\pm$ 13.7&5.50$_{-1.32}^{+1.67}$&16.8$_{-11.0}^{+6.43}$&-\\ 
     SDSS J1343+4155&2.31 $\pm$ 1.046&3.34$_{-1.11}^{+1.38}$&8.13$_{-6.87}^{+4.76}$&6.60 $\pm$ 3.20\\ 
     SDSS J1439+3250&12.4 $\pm$ 6.26&-&-&4.73 $\pm$ 2.84\\  \tableline
     \end{tabular}
\caption{$M_{200}$ values from other papers for several of our systems.  The other papers are \citet{Oguri12}, \citet{Bayliss11} and \citet{Drabek12}.  Note that the values from \citet{Oguri12} have been converted from $M_{vir}$ to $M_{200}$ using the process detailed in the appendix of \citet{Johnston07}.  To convert the errors, we simply converted the upper and lower errors on $M_{vir}$ given in \citet{Oguri12} using the same method.  All $M_{200}$ values have the units $10^{14} h^{-1} M_\sun$.  Though there are significant differences in the $M_{200}$ values for some clusters, overall our values seem consistent with those of the other groups.  See Figure \ref{mass}.} \label{masscomp}
\end{table}
\end{center}

\begin{center}
\begin{table}
\small
\begin{tabular}{c c c c}\tableline \tableline
\bf{System} & \bf{$\theta_E$(arcsec)(this paper)} & \bf{$\theta_E$(arcsec)(West et al.)}&\bf{$\theta_E$(arcsec)(Oguri et al.)}\\ \tableline
SDSS J0900+2234&8.0 $\pm$ 2.7&8.32&-\\
SDSS J0901+1814&6.9 $\pm$ 2.3&6.35&-\\
SDSS J0957+0509&8.2 $\pm$ 2.7&-&$5.2_{-0.5}^{+0.5}$ \\\
SDSS J1038+4849&8.6 $\pm$ 2.9&-&$12.6_{-1.6}^{+1.3}$\\
SDSS J1209+2640&11 $\pm$ 3.7&-&$8.8_{-0.9}^{+0.9}$\\
SDSS J1343+4155&13 $\pm$ 4.3&7.05&$5.4_{-1.6}^{+2.5}$\\ \tableline
     \end{tabular}
    \caption{A comparison of values for Einstein radius measured in this paper, in \citet{West12} and in \citet{Oguri12}.}\label{Einscompare}
 \end{table}
\end{center}

\begin{center}
\begin{table}
\small
\begin{tabular}{c  c  c  c  c}\tableline \tableline
\bf{System}&\bf{$\theta_E$~(arcsec)}&\bf{$M_{lens}~(10^{12}h^{-1}M_\sun$)}&\bf{$\sigma_v~(km ~s^{-1}$)}&\bf{$\theta_E$~(rescaled)}\\ \tableline
SDSS J0900+2234&8.0 $\pm$ 2.7&11 $\pm$ 7.3&648 $\pm$ 108&7.9 $\pm$ 2.7 \\ 
SDSS J0901+1814&6.9 $\pm$ 2.3&5.5 $\pm$ 3.7&564 $\pm$ 93.9&5.9 $\pm$ 2.0\\
SDSS J0957+0509&8.2 $\pm$ 2.7&12 $\pm$ 8.0&680 $\pm$ 113&8.5 $\pm$ 2.8\\ 
SDSS J1038+4849& 8.6 $\pm$ 2.9&15 $\pm$ 10.0&780 $\pm$ 130&11 $\pm$ 3.8 \\
SDSS J1209+2640&11 $\pm$ 3.7&36 $\pm$ 24.0&691 $\pm$ 115&19 $\pm$ 6.2\\ 
SDSS J1318+3942&9.1 $\pm$ 3.0&12 $\pm$ 8.0&336 $\pm$ 55.9&8.2 $\pm$ 2.7\\
SDSS J1343+4155&13 $\pm$ 4.3&24 $\pm$ 16.0&804 $\pm$ 134&12 $\pm$ 3.9\\ 
SDSS J1439+3250\tablenotemark{a}&7.4 $\pm$ 2.5&7.4 $\pm$ 4.9 -10.0 $\pm$ 6.7&596 $\pm$ 99.2 - 708 $\pm$ 118&7.1 $\pm$ 2.4\\
SDSS J1511+4713&5.4 $\pm$ 1.8&6.3 $\pm$ 4.2&631 $\pm$ 105&7.3 $\pm$ 2.4\\
SDSS J1537+6556&8.5 $\pm$ 2.8&8.7 $\pm$ 5.8&715 $\pm$ 119&9.4 $\pm$ 3.1\\ \tableline
     \end{tabular}
    \caption{A summary of the properties measured for the ten strong lensing systems.  Rescaled Einstein radii are Einstein radii projected to fiducial redshifts, of $z_{lens}=0.433$ and $z_{source}=1.65$.  See Equation \ref{fiducial}.}\label{lenses}
    \tablenotetext{a}{Source redshift has not yet been determined for the arc in this system and we can only present a range of redshifts, leading to a range of values for mass and velocity dispersion.}
\end{table}
\end{center}

\clearpage

\clearpage

\clearpage

\clearpage

\end{document}